\documentclass[twocolumn, letterpaper,10pt]{IEEEtran}

\usepackage{cite}

\usepackage{graphicx}
\usepackage{xcolor}
\usepackage{tikz}
\usetikzlibrary{arrows,positioning,shapes.geometric}
\usepackage{pgfplots}
\usepackage{multirow}
\usepackage{upgreek}
\usepackage{amssymb}
\usepackage{amsmath}
\usepackage{cases}
\usepackage{pifont}
\usepackage{bm}
\usepackage{amsthm}

\usepackage{tabularx}
\usepackage{booktabs}
\usepackage{filecontents}
\usepackage{subcaption}
\newcolumntype{Y}{>{\centering\arraybackslash}X}

\usepackage{algorithmic}

\usepackage{array}
\usepackage{url}

\usepackage{circuitikz}
\usetikzlibrary{arrows,shapes.gates.logic.US,shapes.gates.logic.IEC,calc}
\usepackage{hyperref}
\hypersetup{%
	colorlinks=black, 
	breaklinks=false, 
	urlcolor=blue, 
	linkcolor=black, 
	citecolor=blue, 
} 

\newcommand{\fixme}[2]{\ifx&#2&{\leavevmode\color{red}#1}\else{\leavevmode\color{red}FIXME\{}#1{\leavevmode\color{red}\}}\footnote{{\leavevmode\color{red}#2}}\PackageWarning{Fixme}{#1: #2}\fi}

\newcommand{\newstuff}[2]{\ifx&#2&{\leavevmode\color{blue}#1}\else{\leavevmode\color{blue}FIXME\{}#1{\leavevmode\color{blue}\}}\footnote{{\leavevmode\color{blue}#2}}\PackageWarning{Newstuff}{#1: #2}\fi}

\DeclareMathOperator*{\arctanh}{arctanh}

\title{A Multi-Kernel Multi-Code Polar Decoder Architecture}

\author{Gabriele~Coppolino,
	Carlo~Condo,~\IEEEmembership{Member,~IEEE},
        Guido~Masera,~\IEEEmembership{Senior~Member,~IEEE},
        Warren~J.~Gross,~\IEEEmembership{Senior~Member,~IEEE}
\thanks{G.~Coppolino and G.~Masera are with the Department of Electrical and Telecommunications Engineering, Politecnico di Torino, Torino, Italy. e-mail: gabriele.coppolino@studenti.polito.it, guido.masera@polito.it. 
C.~Condo, and W.~J.~Gross are with the Department of Electrical and Computer Engineering, McGill University, Montr\'eal, Qu\'ebec, Canada. e-mail: carlo.condo@mcgill.ca, warren.gross@mcgill.ca.}
}

\begin{document}

\maketitle

\begin{abstract}

Polar codes have received increasing attention in the past decade, and have been selected for the next generation of wireless communication standard. Most research on polar codes has focused on codes constructed from a $2\times2$ polarization matrix, called binary kernel: codes constructed from binary kernels have code lengths that are bound to powers of $2$.
A few recent works have proposed construction methods based on multiple kernels of different dimensions, not only binary ones, allowing code lengths different from powers of $2$. 
In this work, we design and implement the first multi-kernel successive cancellation polar code decoder in literature. It can decode any code constructed with binary and ternary kernels: the architecture, sized for a maximum code length $N_{max}$, is fully flexible in terms of code length, code rate and kernel sequence. The decoder can achieve frequency of more than $1$ GHz in $65$ nm CMOS technology, and a throughput of $615$ Mb/s. The area occupation ranges between $0.11$ mm$^2$ for $N_{max}=256$ and $2.01$ mm$^2$ for $N_{max}=4096$. Implementation results show an unprecedented degree of flexibility: with $N_{max}=4096$, up to $55$ code lengths can be decoded with the same hardware, along with any kernel sequence and code rate.

\end{abstract}


\section{Introduction} \label{sec:intro}

Polar codes are capacity-achieving error correcting codes, characterized by a low-complexity encoding and decoding process \cite{arikan}. They have been chosen to be adopted in the fifth generation of wireless communication standards (5G) \cite{3gpp_polar}, that foresees a variety of scenarios. Thus, coding schemes targeting low latency, low power, and high performance must be devised. Error correction performance and decoding speed are heavily influenced by the polar code block length, and the different scenarios demand a wide range of code lengths. 

The majority of current research is focused on polar codes recursively constructed from a $2\times 2$ polarization matrix, also called a binary kernel \cite{arikan}. The code lengths of polar codes constructed from binary kernels are bound to powers of $2$. This is a strong limitation, that is currently overcome with rate-matching schemes \cite{Chandesris_ICC17,Bioglio_punct}, whose performance and optimality is hard to evaluate a priori. A few recent works have proposed construction methods based on multiple kernels of different dimensions \cite{Presman_TIT15,Gabry_MK,Bioglio_MK}. Multi-kernel polar codes can have block lengths different from powers of $2$, at the cost of more complex decoding algorithm update rules. In \cite{Gabry_MK}, it has been shown that multi-kernel codes can outperform codes of the same length obtained through the application of state-of-the-art puncturing and shortening schemes. At a frame error rate (FER) of almost $10^{-3}$, multi-kernel codes yield gains ranging from $0.1$~dB to $1.1$~dB.

Polar code decoder architectures in literature focus mainly on design-time flexibility
\cite{leroux,yuan_2bitDecoding,Fan_Tsui}, with parametrized designs that can be implemented to decode a particular code. Some decoders guarantee code-rate online flexibility \cite{hashemi_SSCL_TCASI,hashemi_FSSCL_TSP,Dizdar_TCASI,Che_ICASSP16}: while the decoder can decode a single code length, any code rate is supported with the same hardware. The decoder architectures presented in \cite{sarkis,Ercan_SIPS17} target binary kernels only, and are online flexible in terms of both code rate and code length. However, a different decoding program must be stored for every considered combination of code length and rate, leading to huge area occupation. The unrolled architecture presented in \cite{giard_unrolled} can decode a small set of binary nested code lengths and rates.

In this work, we consider multi-kernel polar codes constructed from binary and ternary ($3\times 3$) kernels, and we propose a flexible decoder architecture. The presented design can decode any code constructed from any combination of binary and ternary kernels, up to a maximum code length defined at design time, and any code rate. It is the first multi-kernel decoder in literature, yielding an unmatched degree of flexibility, with up to 55 supported code lengths in the considered case study. Implementation results in $65$~nm CMOS technology show an achievable frequency of more that $1$ GHz and $615$ Mbps coded throughput.

The remainder of the paper is organized as follows. In Section \ref{sec:prel}, we introduce polar codes construction and decoding, while in Section \ref{sec:codes} we show the error-correction performance of some multi-kernel codes. Section \ref{sec:HW} details the proposed decoder architecture, while implementation results are given in \ref{sec:impl}, together with a comparison with the state of the art. Conclusions are drawn in Section \ref{sec:conc}.

\section{Preliminaries} \label{sec:prel}
\subsection{Polar Codes}
A polar code $\mathcal{P}(N,K)$ is a linear block code of length $N$ and rate $K/N$, that relies on a phenomenon called channel polarization \cite{arikan}. When $N$ tends to infinite, the symmetric capacity of each bit-channel tends towards either $0$ or $1$, thus identifying very reliable and very unreliable channels.

Let us assume $N=2^n$, where $n\geq1$, and let $\mathbf{u} = (u_0, u_1, \ldots, u_{N-1})$ be the $N$-bit vector input to the encoder.
The $K$ information bits are assigned to the $K$ most reliable channels of $\mathbf{u}$, while the remaining $N-K$ are fixed to a known value (usually $0$), and are known as frozen bits. The ensemble of their indices is the frozen set $\mathcal{F}$.

\begin{figure}[t!]
	\resizebox {\columnwidth} {!} {
		\begin{tikzpicture}
	\node at (-2,-.25) {$u_0$};
	\node at (-2,-1.25) {$u_1$};
	\node at (-2,-2.25) {$u_2$};
	\node at (-2,-3.25) {$u_3$};
	\node at (-2,-4.25) {$u_4$};
	\node at (-2,-5.25) {$u_5$};
	\node at (-2,-6.25) {$u_6$};
	\node at (-2,-7.25) {$u_7$};
	
	\draw (-1.5,-.25) -- (0,-.25);
	\draw (-1.5,-1.25) -- (0,-1.25);
	\draw (-1.5,-2.25) -- (0,-2.25);
	\draw (-1.5,-3.25) -- (0,-3.25);
	\draw (-1.5,-4.25) -- (0,-4.25);
	\draw (-1.5,-5.25) -- (0,-5.25);
	\draw (-1.5,-6.25) -- (0,-6.25);
	\draw (-1.5,-7.25) -- (0,-7.25);
	
	\node at (0.5,-8) {Stage 3};
	\draw [dashed, very thick, gray!50] (0,0) rectangle (1,-1.5);
	\draw [dashed, very thick, gray!50] (0,-2) rectangle (1,-3.5);
	\draw [dashed, very thick, gray!50] (0,-4) rectangle (1,-5.5);
	\draw [dashed, very thick, gray!50] (0,-6) rectangle (1,-7.5);
	
	
	\draw (1,-.25) -- (2,-.25);
	\draw (1,-1.25) -- (2,-1.25);
	\draw (1,-2.25) -- (2,-2.25);
	\draw (1,-3.25) -- (2,-3.25);
	\draw (1,-4.25) -- (2,-4.25);
	\draw (1,-5.25) -- (2,-5.25);
	\draw (1,-6.25) -- (2,-6.25);
	\draw (1,-7.25) -- (2,-7.25);
	
	\draw[gray!60] (2,0) rectangle (3,-7.5);
	\node at (2.5,-8) {$P_3$};
	\draw (2,-.25) -- (3,-.25);
	\draw (2,-1.25) -- (3,-4.25);
	\draw (2,-2.25) -- (3,-1.25);
	\draw (2,-3.25) -- (3,-5.25);
	\draw (2,-4.25) -- (3,-2.25);
	\draw (2,-5.25) -- (3,-6.25);
	\draw (2,-6.25) -- (3,-3.25);
	\draw (2,-7.25) -- (3,-7.25);
	
	\draw (3,-.25) -- (4,-.25);
	\draw (3,-1.25) -- (4,-1.25);
	\draw (3,-2.25) -- (4,-2.25);
	\draw (3,-3.25) -- (4,-3.25);
	\draw (3,-4.25) -- (4,-4.25);
	\draw (3,-5.25) -- (4,-5.25);
	\draw (3,-6.25) -- (4,-6.25);
	\draw (3,-7.25) -- (4,-7.25);
	
	\node at (4.5,-8) {Stage 2};
	\draw [dashed, very thick, gray!50] (4,0) rectangle (5,-1.5);
	\draw [dashed, very thick, gray!50] (4,-2) rectangle (5,-3.5);
	\draw [dashed, very thick, gray!50] (4,-4) rectangle (5,-5.5);
	\draw [dashed, very thick, gray!50] (4,-6) rectangle (5,-7.5);
	
	
	\draw (5,-.25) -- (6,-.25);
	\draw (5,-1.25) -- (6,-1.25);
	\draw (5,-2.25) -- (6,-2.25);
	\draw (5,-3.25) -- (6,-3.25);
	\draw (5,-4.25) -- (6,-4.25);
	\draw (5,-5.25) -- (6,-5.25);
	\draw (5,-6.25) -- (6,-6.25);
	\draw (5,-7.25) -- (6,-7.25);
	
	\draw[gray!60] (6,0) rectangle (7,-7.5);	
	\node at (6.5,-8) {$P_2$};
	\draw (6,-.25) -- (7,-.25);
	\draw (6,-1.25) -- (7,-2.25);
	\draw (6,-2.25) -- (7,-1.25);
	\draw (6,-3.25) -- (7,-3.25);
	\draw (6,-4.25) -- (7,-4.25);
	\draw (6,-5.25) -- (7,-6.25);
	\draw (6,-6.25) -- (7,-5.25);
	\draw (6,-7.25) -- (7,-7.25);
	
	\draw (7,-.25) -- (8,-.25);
	\draw (7,-1.25) -- (8,-1.25);
	\draw (7,-2.25) -- (8,-2.25);
	\draw (7,-3.25) -- (8,-3.25);
	\draw (7,-4.25) -- (8,-4.25);
	\draw (7,-5.25) -- (8,-5.25);
	\draw (7,-6.25) -- (8,-6.25);
	\draw (7,-7.25) -- (8,-7.25);
	
	\node at (8.5,-8) {Stage 1};
	\draw [dashed, very thick, gray!50] (8,0) rectangle (9,-1.5);
	\draw [dashed, very thick, gray!50] (8,-2) rectangle (9,-3.5);
	\draw [dashed, very thick, gray!50] (8,-4) rectangle (9,-5.5);
	\draw [dashed, very thick, gray!50] (8,-6) rectangle (9,-7.5);
	
	
	\draw (9,-.25) -- (10,-.25);
	\draw (9,-1.25) -- (10,-1.25);
	\draw (9,-2.25) -- (10,-2.25);
	\draw (9,-3.25) -- (10,-3.25);
	\draw (9,-4.25) -- (10,-4.25);
	\draw (9,-5.25) -- (10,-5.25);
	\draw (9,-6.25) -- (10,-6.25);
	\draw (9,-7.25) -- (10,-7.25);
	
	\draw[gray!60] (10,0) rectangle (11,-7.5);
	\node at (10.5,-8) {$P_1$};	
	\draw (10,-.25) -- (11,-.25);
	\draw (10,-1.25) -- (11,-4.25);
	\draw (10,-2.25) -- (11,-2.25);
	\draw (10,-3.25) -- (11,-6.25);
	\draw (10,-4.25) -- (11,-1.25);
	\draw (10,-5.25) -- (11,-5.25);
	\draw (10,-6.25) -- (11,-3.25);
	\draw (10,-7.25) -- (11,-7.25);
	
	\draw [->] (11,-.25) -- (12.5,-.25);
	\draw [->] (11,-1.25) -- (12.5,-1.25);
	\draw [->] (11,-2.25) -- (12.5,-2.25);
	\draw [->] (11,-3.25) -- (12.5,-3.25);
	\draw [->] (11,-4.25) -- (12.5,-4.25);
	\draw [->] (11,-5.25) -- (12.5,-5.25);
	\draw [->] (11,-6.25) -- (12.5,-6.25);
	\draw [->] (11,-7.25) -- (12.5,-7.25);
	
	\node at (13,-.25) {$x_0$};
	\node at (13,-1.25) {$x_1$};
	\node at (13,-2.25) {$x_2$};
	\node at (13,-3.25) {$x_3$};
	\node at (13,-4.25) {$x_4$};
	\node at (13,-5.25) {$x_5$};
	\node at (13,-6.25) {$x_6$};
	\node at (13,-7.25) {$x_7$};
	
	\foreach \x in {0,4,8}{
	\foreach \t in {-.25,-2.25,-4.25,-6.25}{
	\draw [->](\x,\t) -- (\x+.4,\t);
	\draw (\x+.6,\t) -- (\x+1,\t);
	\draw (\x+.5,\t) circle [radius=.1];
	\draw (\x+.5,\t-.1) -- (\x+.5,\t+.1);
	\draw (\x+.5-.1,\t) -- (\x+.5+.1,\t);
	\draw[->] (\x+.5,\t-1) -- (\x+.5,\t-.1);
	\fill (\x+.5,\t-1) circle[radius=.05];
	\draw (\x,\t-1) -- (\x+1,\t-1);
	}}

	\end{tikzpicture}
}
	\caption{Tanner graph for a $N=8$ polar code.}
	\label{fig:sez_prel:n8_tanner_graph}
\end{figure}
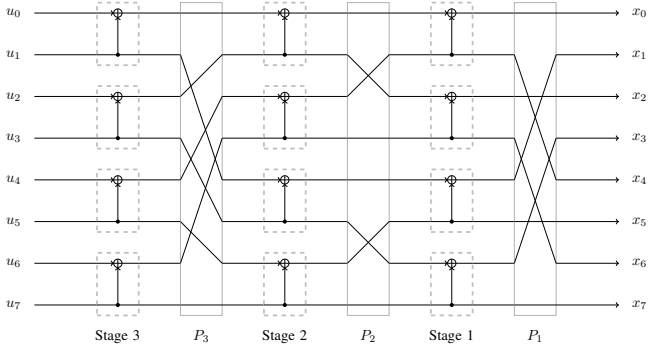

The encoding process can be represented through the linear transformation $\mathbf{x} = \mathbf{u}\mathbf{G}$, where $\mathbf{G}=\mathbf{T_2}^{\otimes n}$ is the generator matrix, expressed through the \textit{n}-th Kronecker product of the matrix $\mathbf{T_2}$. The matrix $\mathbf{T_2}$ is a binary polarization matrix, or kernel, defined as follows:
\begin{equation*}
\mathbf{T_2}=\left[\begin{matrix} 1 & 0 \\ 1 & 1\end{matrix}\right].
\end{equation*}
From the definition of $\mathbf{G}$, the recursive nature of the encoding process can be noticed: a polar code of length $N$ can in fact be obtained as the concatenation of two $N/2$ polar codes. Polar code encoding can also be portrayed through a Tanner graph, as shown in Fig. \ref{fig:sez_prel:n8_tanner_graph} for an $N=8$ code. Each stage depicts a Kronecker product, and the dashed boxes represent each $\mathbf{T_2}$ operation. Between neighbouring stages permutations are inserted, in which the bit-indices of the inputs are cyclically rotated to the right by one place \cite{arikan}.

\subsection{Successive Cancellation Decoding}

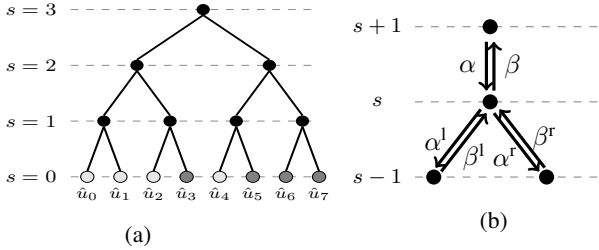
\begin{figure}[t!]
  \centering
\begin{subfigure}{0.45\columnwidth}
	\resizebox{1.2\columnwidth} {2.8cm}{
  \begin{tikzpicture}[scale=1.8, thick]

\draw [very thin,gray,dashed] (-2,-.5) -- (-.125,-.5);
\draw [very thin,gray,dashed] (-2,-1) -- (-.125,-1);
\draw [very thin,gray,dashed] (-2,-1.5) -- (-.125,-1.5);
\draw [very thin,gray,dashed] (-2,-2) -- (-.125,-2);


\fill (-1,-.5) circle [radius=.05];

\fill (-1.5,-1) circle [radius=.05];
\fill (-.5,-1) circle [radius=.05];

\fill (-1.75,-1.5) circle [radius=.05];
\fill (-1.25,-1.5) circle [radius=.05];
\fill (-.75,-1.5) circle [radius=.05];
\fill (-.25,-1.5) circle [radius=.05];

\fill [gray!20](-1.875,-2) circle [radius=.05];
\fill [gray!20](-1.625,-2) circle [radius=.05];
\fill [gray!20](-1.375,-2) circle [radius=.05];
\fill [gray](-1.125,-2) circle [radius=.05];
\fill [gray!20](-.875,-2) circle [radius=.05];
\fill [gray](-.625,-2) circle [radius=.05];
\fill [gray](-.375,-2) circle [radius=.05];
\fill [gray](-.125,-2) circle [radius=.05];
\draw [very thin]  (-1.875,-2) circle [radius=.05];
\draw [very thin] (-1.625,-2) circle [radius=.05];
\draw [very thin] (-1.375,-2) circle [radius=.05];
\draw [very thin] (-1.125,-2) circle [radius=.05];
\draw [very thin] (-.875,-2) circle [radius=.05];
\draw [very thin] (-.625,-2) circle [radius=.05];
\draw [very thin] (-.375,-2) circle [radius=.05];
\draw [very thin] (-.125,-2) circle [radius=.05];

\node at (-1.875,-2-.15) {\scriptsize $\hat{u}_{0}$};
\node at (-1.625,-2-.15) {\scriptsize $\hat{u}_1$};
\node at (-1.375,-2-.15) {\scriptsize $\hat{u}_2$};
\node at (-1.125,-2-.15) {\scriptsize $\hat{u}_3$};
\node at (-.875,-2-.15)  {\scriptsize $\hat{u}_4$};
\node at (-.625,-2-.15)  {\scriptsize $\hat{u}_5$};
\node at (-.375,-2-.15) {\scriptsize $\hat{u}_6$};
\node at (-.125,-2-.15)  {\scriptsize $\hat{u}_7$};


\draw (-1,-.55) -- (-1.5,-.95);
\draw (-1,-.55) -- (-.5,-.95);

\draw (-1.5,-1.05) -- (-1.75,-1.45);
\draw (-1.5,-1.05) -- (-1.25,-1.45);
\draw (-.5,-1.05) -- (-.75,-1.45);
\draw (-.5,-1.05) -- (-.25,-1.45);

\draw (-1.75,-1.55) -- (-1.875,-1.95);
\draw (-1.75,-1.55) -- (-1.625,-1.95);
\draw (-1.25,-1.55) -- (-1.375,-1.95);
\draw (-1.25,-1.55) -- (-1.125,-1.95);
\draw (-.75,-1.55) -- (-.875,-1.95);
\draw (-.75,-1.55) -- (-.625,-1.95);
\draw (-.25,-1.55) -- (-.375,-1.95);
\draw (-.25,-1.55) -- (-.125,-1.95);

\node at (-2.3,-.5) {\footnotesize $s=3$};
\node at (-2.3,-1) {\footnotesize $s=2$};
\node at (-2.3,-1.5) {\footnotesize $s=1$};
\node at (-2.3,-2) {\footnotesize $s=0$};

\end{tikzpicture}}
  \caption{}
  \label{fig:sez_prel:n8_decoding_tree}
\end{subfigure}
\hspace{0.5cm}
\begin{subfigure}{0.45\columnwidth}
  \begin{tikzpicture}[scale=.5]

\draw [very thin,gray,dashed] (-2,0) -- (3,0);
\draw [very thin,gray,dashed] (-2,-2) -- (3,-2);
\draw [very thin,gray,dashed] (-2,-4) -- (3,-4);

\node at (-3,0) {\footnotesize $s+1$};
\node at (-3,-2) {\footnotesize $s$};
\node at (-3,-4) {\footnotesize $s-1$};

\fill (0,0) circle [radius=.2];
\fill (0,-2) circle [radius=.2];
\fill (-1.5,-4) circle [radius=.2];
\fill (1.5,-4) circle [radius=.2];

\draw [->,very thick] (-.1,-.4) -- (-.1,-1.7) node [left,midway,rotate=0] {$\alpha$};
\draw [->,very thick] (.1,-1.7) -- (.1,-.4) node [right,midway,rotate=0] {$\beta$};

\draw [->,very thick] (-.25,-2.2) -- (-1.45,-3.75) node [left,midway,rotate=0] {$\alpha^{\text{l}}$};
\draw [->,very thick] (-1.3,-3.85) -- (-.1,-2.3) node [right,near start,rotate=0] {$\beta^{\text{l}}$};
\draw [<-,very thick] (.25,-2.2) -- (1.45,-3.75) node [right,midway,rotate=0] {$\beta^{\text{r}}$};
\draw [<-,very thick] (1.3,-3.85) -- (.1,-2.3) node [left,near start,rotate=0] {$\alpha^{\text{r}}$};

\end{tikzpicture}
  \caption{}
  \label{fig:sez_prel:bin_message_passing}
\end{subfigure}
\caption{(a) Decoding tree for a $\mathcal{P}(8,4)$ polar code and (b) binary node message passing.}
  \label{fig:sez_prel:n8tree_messagePassing}
\end{figure}

In \cite{arikan} a first successive cancellation (SC) decoding algorithm has been proposed. It can be represented as a binary search tree where all the nodes must be explored, with priority being given to left branches. An example of a $\mathcal{P}(8,4)$ polar code SC decoding tree is shown in Fig. \ref{fig:sez_prel:n8_decoding_tree}: the leaf nodes at stage $s=0$ can be either information bits (dark gray) or frozen bits (light gray). 

Let us call $\mathbf{y} = (y_0, y_1, \ldots, y_{N-1})$ the vector of logarithmic likelihood ratios (LLRs) obtained at the channel output, and $\mathbf{\hat{u}}$ the estimated vector output by the decoder.
The decoding starts from the root node, and at each node information is passed from parent to child according to the scheme shown in Fig. \ref{fig:sez_prel:bin_message_passing}. The LLR value $\alpha$ is received and used to compute $\alpha^l$, then $\beta^l$ is obtained and used to compute $\alpha^r$. Once $\beta^r$ is available, $\beta$ can be computed. Once a leaf node is reached, the value of $\hat{u}_i$ is estimated. If index $i\in {\mathcal{F}}$, its value is set to $0$, otherwise a hard decision on the sign of $\alpha$ is performed. Calling $N_s$ the length of the polar code at stage $s$, we can define $\forall i \in \left(0,1,...,\frac{N_s}{2}-1\right)$: 

\begin{multline}
\alpha^l_i= 2\arctanh \left(\tanh \frac{\alpha_i}{2} \cdot \tanh \frac{\alpha_{i+\frac{N_s}{2}}}{2} \right)\\
\simeq \varphi \left(\alpha_i\right) \varphi \left(\alpha_{i+\frac{N_s}{2}}\right) \min\left(\left\rvert \alpha_i \right\rvert, \left\rvert \alpha_{i+\frac{N_s}{2}}\right\rvert\right)~,
\label{eqn:sez_prel:f_bin} 
\end{multline}
\begin{align}
\alpha^r_i=\left(1-2 \beta^l_i\right)\alpha_i  + \alpha_{i+\frac{N_s}{2}}~,
\label{eqn:sez_prel:g_bin}
\end{align}
\begin{align}
\left[\beta_i, \beta_{i+\frac{N_s}{2}}\right] = \left[ \beta^l_i \oplus \beta^r_i, \beta^r_i \right]~,
\label{eqn:sez_prel:comb_bin}
\end{align}
where $\oplus$ represents the XOR operation and $\varphi()$ is a function returning the sign of the argument. In (\ref{eqn:sez_prel:f_bin}), both the exact and the approximate (hardware-friendly) computation, proposed in \cite{leroux}, are shown. At leaf nodes, $\beta$ is initialized as $\hat{u}_i$ (\ref{eqn:sez_prel:HD}), where $i$ is the index identifying the current leaf node.

\begin{equation}
\hat{u}_i =
\begin{cases}
0 & 
\mbox{if $\alpha\geq 0$ or $i\in \mathcal{F}$}\\
1 &  \mbox{otherwise}
\end{cases}
\label{eqn:sez_prel:HD}
\end{equation}

\subsection{Multi-kernel construction}
\begin{figure}[t!]
	\resizebox {\columnwidth} {!} {
			\begin{tikzpicture}
	\node at (-2,-.25) {$u_0$};
	\node at (-2,-1.25) {$u_1$};
	\node at (-2,-2.25) {$u_2$};
	\node at (-2,-3.25) {$u_3$};
	\node at (-2,-4.25) {$u_4$};
	\node at (-2,-5.25) {$u_5$};
	\node at (-2,-6.25) {$u_6$};
	\node at (-2,-7.25) {$u_7$};
	\node at (-2,-8.25) {$u_8$};
	\node at (-2,-9.25) {$u_9$};
	\node at (-2,-10.25) {$u_{10}$};
	\node at (-2,-11.25) {$u_{11}$};
	
	\draw (-1.5,-.25) -- (0,-.25);
	\draw (-1.5,-1.25) -- (0,-1.25);
	\draw (-1.5,-2.25) -- (0,-2.25);
	\draw (-1.5,-3.25) -- (0,-3.25);
	\draw (-1.5,-4.25) -- (0,-4.25);
	\draw (-1.5,-5.25) -- (0,-5.25);
	\draw (-1.5,-6.25) -- (0,-6.25);
	\draw (-1.5,-7.25) -- (0,-7.25);
	\draw (-1.5,-8.25) -- (0,-8.25);
	\draw (-1.5,-9.25) -- (0,-9.25);
	\draw (-1.5,-10.25) -- (0,-10.25);
	\draw (-1.5,-11.25) -- (0,-11.25);
	
	\node at (0.5,-12) {Stage 3};
	\draw [dashed, very thick, gray!50] (0,0) rectangle (1,-1.5);
	\draw [dashed, very thick, gray!50] (0,-2) rectangle (1,-3.5);
	\draw [dashed, very thick, gray!50] (0,-4) rectangle (1,-5.5);
	\draw [dashed, very thick, gray!50] (0,-6) rectangle (1,-7.5);
	\draw [dashed, very thick, gray!50] (0,-8) rectangle (1,-9.5);
	\draw [dashed, very thick, gray!50] (0,-10) rectangle (1,-11.5);
	
	
	\draw (1,-.25) -- (2,-.25);
	\draw (1,-1.25) -- (2,-1.25);
	\draw (1,-2.25) -- (2,-2.25);
	\draw (1,-3.25) -- (2,-3.25);
	\draw (1,-4.25) -- (2,-4.25);
	\draw (1,-5.25) -- (2,-5.25);
	\draw (1,-6.25) -- (2,-6.25);
	\draw (1,-7.25) -- (2,-7.25);
	\draw (1,-8.25) -- (2,-8.25);
	\draw (1,-9.25) -- (2,-9.25);
	\draw (1,-10.25) -- (2,-10.25);
	\draw (1,-11.25) -- (2,-11.25);
	
	\draw[gray!60] (2,0) rectangle (3,-11.5);
	\node at (2.5,-12) {$P_3$};
	\draw (2,-.25) -- (3,-.25);
	\draw (2,-1.25) -- (3,-2.25);
	\draw (2,-2.25) -- (3,-4.25);
	\draw (2,-3.25) -- (3,-6.25);
	\draw (2,-4.25) -- (3,-8.25);
	\draw (2,-5.25) -- (3,-10.25);
	\draw (2,-6.25) -- (3,-1.25);
	\draw (2,-7.25) -- (3,-3.25);
	\draw (2,-8.25) -- (3,-5.25);
	\draw (2,-9.25) -- (3,-7.25);
	\draw (2,-10.25) -- (3,-9.25);
	\draw (2,-11.25) -- (3,-11.25);
	
	\draw (3,-.25) -- (4,-.25);
	\draw (3,-1.25) -- (4,-1.25);
	\draw (3,-2.25) -- (4,-2.25);
	\draw (3,-3.25) -- (4,-3.25);
	\draw (3,-4.25) -- (4,-4.25);
	\draw (3,-5.25) -- (4,-5.25);
	\draw (3,-6.25) -- (4,-6.25);
	\draw (3,-7.25) -- (4,-7.25);
	\draw (3,-8.25) -- (4,-8.25);
	\draw (3,-9.25) -- (4,-9.25);
	\draw (3,-10.25) -- (4,-10.25);
	\draw (3,-11.25) -- (4,-11.25);
	
	\node at (4.5,-12) {Stage 2};
	\draw [dashed, very thick, gray!50] (4,0) rectangle (5,-2.5);
	\draw [dashed, very thick, gray!50] (4,-3) rectangle (5,-5.5);
	\draw [dashed, very thick, gray!50] (4,-6) rectangle (5,-8.5);
	\draw [dashed, very thick, gray!50] (4,-9) rectangle (5,-11.5);
	
	
	\draw (5,-.25) -- (6,-.25);
	\draw (5,-1.25) -- (6,-1.25);
	\draw (5,-2.25) -- (6,-2.25);
	\draw (5,-3.25) -- (6,-3.25);
	\draw (5,-4.25) -- (6,-4.25);
	\draw (5,-5.25) -- (6,-5.25);
	\draw (5,-6.25) -- (6,-6.25);
	\draw (5,-7.25) -- (6,-7.25);
	\draw (5,-8.25) -- (6,-8.25);
	\draw (5,-9.25) -- (6,-9.25);
	\draw (5,-10.25) -- (6,-10.25);
	\draw (5,-11.25) -- (6,-11.25);
	
	\draw[gray!60] (6,0) rectangle (7,-11.5);	
	\node at (6.5,-12) {$P_2$};
	\draw (6,-.25) -- (7,-.25);
	\draw (6,-1.25) -- (7,-3.25);
	\draw (6,-2.25) -- (7,-1.25);
	\draw (6,-3.25) -- (7,-4.25);
	\draw (6,-4.25) -- (7,-2.25);
	\draw (6,-5.25) -- (7,-5.25);
	\draw (6,-6.25) -- (7,-6.25);
	\draw (6,-7.25) -- (7,-9.25);
	\draw (6,-8.25) -- (7,-7.25);
	\draw (6,-9.25) -- (7,-10.25);
	\draw (6,-10.25) -- (7,-8.25);
	\draw (6,-11.25) -- (7,-11.25);
	
	\draw (7,-.25) -- (8,-.25);
	\draw (7,-1.25) -- (8,-1.25);
	\draw (7,-2.25) -- (8,-2.25);
	\draw (7,-3.25) -- (8,-3.25);
	\draw (7,-4.25) -- (8,-4.25);
	\draw (7,-5.25) -- (8,-5.25);
	\draw (7,-6.25) -- (8,-6.25);
	\draw (7,-7.25) -- (8,-7.25);
	\draw (7,-8.25) -- (8,-8.25);
	\draw (7,-9.25) -- (8,-9.25);
	\draw (7,-10.25) -- (8,-10.25);
	\draw (7,-11.25) -- (8,-11.25);
	
	\node at (8.5,-12) {Stage 1};
	\draw [dashed, very thick, gray!50] (8,0) rectangle (9,-1.5);
	\draw [dashed, very thick, gray!50] (8,-2) rectangle (9,-3.5);
	\draw [dashed, very thick, gray!50] (8,-4) rectangle (9,-5.5);
	\draw [dashed, very thick, gray!50] (8,-6) rectangle (9,-7.5);
	\draw [dashed, very thick, gray!50] (8,-8) rectangle (9,-9.5);
	\draw [dashed, very thick, gray!50] (8,-10) rectangle (9,-11.5);
	
	
	\draw (9,-.25) -- (10,-.25);
	\draw (9,-1.25) -- (10,-1.25);
	\draw (9,-2.25) -- (10,-2.25);
	\draw (9,-3.25) -- (10,-3.25);
	\draw (9,-4.25) -- (10,-4.25);
	\draw (9,-5.25) -- (10,-5.25);
	\draw (9,-6.25) -- (10,-6.25);
	\draw (9,-7.25) -- (10,-7.25);
	\draw (9,-8.25) -- (10,-8.25);
	\draw (9,-9.25) -- (10,-9.25);
	\draw (9,-10.25) -- (10,-10.25);
	\draw (9,-11.25) -- (10,-11.25);
	
	\draw[gray!60] (10,0) rectangle (11,-11.5);
	\node at (10.5,-12) {$P_1$};	
	\draw (10,-.25) -- (11,-.25);
	\draw (10,-1.25) -- (11,-6.25);
	\draw (10,-2.25) -- (11,-2.25);
	\draw (10,-3.25) -- (11,-8.25);
	\draw (10,-4.25) -- (11,-4.25);
	\draw (10,-5.25) -- (11,-10.25);
	\draw (10,-6.25) -- (11,-1.25);
	\draw (10,-7.25) -- (11,-7.25);
	\draw (10,-8.25) -- (11,-3.25);
	\draw (10,-9.25) -- (11,-9.25);
	\draw (10,-10.25) -- (11,-5.25);
	\draw (10,-11.25) -- (11,-11.25);
	
	\draw [->] (11,-.25) -- (12.5,-.25);
	\draw [->] (11,-1.25) -- (12.5,-1.25);
	\draw [->] (11,-2.25) -- (12.5,-2.25);
	\draw [->] (11,-3.25) -- (12.5,-3.25);
	\draw [->] (11,-4.25) -- (12.5,-4.25);
	\draw [->] (11,-5.25) -- (12.5,-5.25);
	\draw [->] (11,-6.25) -- (12.5,-6.25);
	\draw [->] (11,-7.25) -- (12.5,-7.25);
	\draw [->] (11,-8.25) -- (12.5,-8.25);
	\draw [->] (11,-9.25) -- (12.5,-9.25);
	\draw [->] (11,-10.25) -- (12.5,-10.25);
	\draw [->] (11,-11.25) -- (12.5,-11.25);
	
	\node at (13,-.25) {$x_0$};
	\node at (13,-1.25) {$x_1$};
	\node at (13,-2.25) {$x_2$};
	\node at (13,-3.25) {$x_3$};
	\node at (13,-4.25) {$x_4$};
	\node at (13,-5.25) {$x_5$};
	\node at (13,-6.25) {$x_6$};
	\node at (13,-7.25) {$x_7$};
	\node at (13,-8.25) {$x_8$};
	\node at (13,-9.25) {$x_9$};
	\node at (13,-10.25) {$x_{10}$};
	\node at (13,-11.25) {$x_{11}$};
	
	\foreach \c in {0,8}{
	\foreach \x in {0,-2,-4,-6,-8,-10}{
	\draw (\c, \x-1.25)--(\c+1, \x-1.25);
	\draw [->] (\c+.5, \x-1.25)--(\c+.5, \x-.25-.1);
	\draw [->] (\c, \x-.25)--(\c+.4, \x-.25);
	\draw (\c+.6, \x-.25)--(\c+1, \x-.25);
	\draw (\c+.5, \x-.25) circle [radius=.1];
	\draw (\c+.5-.1, \x-.25) -- (\c+.5+.1, \x-.25);
	\draw (\c+.5, \x-.25+.1) -- (\c+.5, \x-.25-.1);		
	\fill (\c+.5, \x-1.25) circle [radius=.05];
	}}

	\foreach \x in {0,-3,-6,-9}{
	\draw [->] (4,\x-2.25) -- (4.1,\x-2.25);
	\draw (4.2,\x-2.25) circle [radius=.1];
	\draw (4.2-.1,\x-2.25) -- (4.2+.1,\x-2.25);
	\draw (4.2,\x-2.25-.1) -- (4.2,\x-2.25+.1);
	\draw [->] (4.3,\x-2.25) -- (4.6,\x-2.25);
	\draw (4.7,\x-2.25) circle [radius=.1];
	\draw (4.7+.1,\x-2.25) -- (4.7-.1,\x-2.25);
	\draw (4.7,\x-2.25+.1) -- (4.7,\x-2.25-.1);
	\draw (4.8,\x-2.25) -- (5,\x-2.25);
	\fill (4.45,\x-2.25) circle [radius=.05];
	\draw (4.45,\x-2.25) -- (4.45,\x-2.25+.5) -- (4.9,\x-2.25+.5) -- (4.9,\x-1.25) -- (5,\x-1.25);
	\draw (4,\x-1.25) -- (4.7,\x-1.25);
	\fill (4.7,\x-1.25) circle [radius=.05];
	\draw [->] (4.7,\x-1.25) -- (4.7,\x-2.25+.1);
	\draw [->] (4.7,\x-1.25) -- (4.7,\x-.25-.1);
	\draw (4.7,\x-.25) circle [radius=.1];
	\draw (4.7+.1,\x-.25) -- (4.7-.1,\x-.25);
	\draw (4.7,\x-.25+.1) -- (4.7,\x-.25-.1);
	\draw [->] (4,\x-.25) -- (4.6,\x-.25);
	\draw (4.8,\x-.25) -- (5,\x-.25);
	\draw [->] (4.2,\x-.25) -- (4.2,\x-2.25+.1);
	\fill (4.2,\x-.25) circle [radius=.05];
	}

	\end{tikzpicture}
	}
	\caption{Tanner graph for a $N=12$ polar code.}
	\label{fig:sez_prel:n12_tanner_graph}
\end{figure}
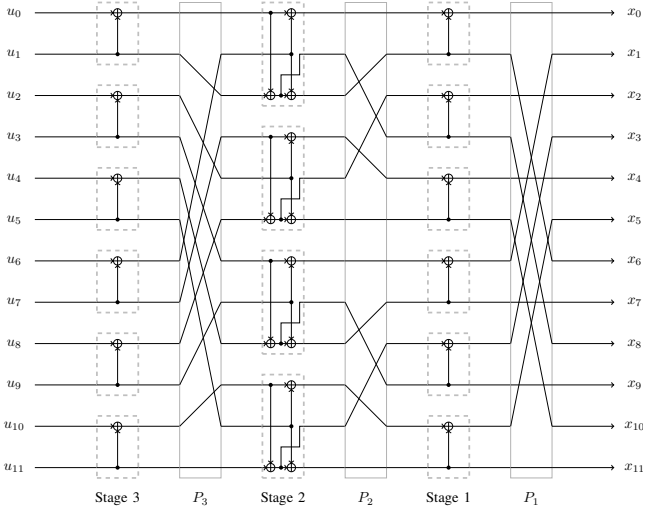

In \cite{Gabry_MK} a generalized construction method for polar codes has been presented: together with $\mathbf{T_2}$, larger kernels have been investigated. Thus, the matrix $\mathbf{G}$ is composed of a series of Kronecker products between kernels of different sizes. 
Ternary kernels, i.e. kernels of dimensions $3\times 3$, have been considered in \cite{Gabry_MK}, where the proposed polarization matrix is 
$$\mathbf{T_3}=\left[\begin{matrix} 1 & 1 & 1 \\ 1 & 0 & 1 \\ 0 & 1 & 1\end{matrix}\right] ~.$$
Fig. \ref{fig:sez_prel:n12_tanner_graph} portrays the Tanner graph for an $N=12$ code constructed with a kernel sequence $\mathbf{T_2} \otimes \mathbf{T_3} \otimes \mathbf{T_2}$. 
As in the binary case, inter-stage permutations are required to reshuffle indices. For each stage $i>1$, the permutation matrix $\mathbf{P_i}$ can be found as
\begin{equation*}
\mathbf{P_i}=(\mathbf{Q_i}| \mathbf{Q_i}+N_{i+1}| \mathbf{Q_i}+2N_{i+1}|\ldots|\mathbf{Q_i}+(N/N_{i+1})N_{i+1})~,
\end{equation*}
where $\mathbf{Q_i}$ is the so-called canonical permutation introduced in \cite{Gabry_MK}, $N_i=\prod_{j=1}^{i-1}n_j$, and $n_j\times n_j$ are the dimensions of the $j$-th kernel of the Kronecker product. Finally, $\mathbf{P_1}$ is computed in order to re-align output indices with those relative to the encoder input, considering all the previous permutations. 

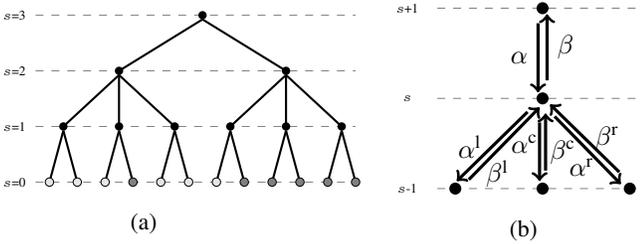
\begin{figure}[t!]
	\centering
	\begin{subfigure}{0.45\columnwidth}
			\begin{tikzpicture}[scale=.37, thick]
	
	\node[left] at (.5,0) {\tiny $s$=0};
	\node[left] at (.5,2) {\tiny $s$=1};
	\node[left] at (.5,4) {\tiny $s$=2};
	\node[left] at (.5,6) {\tiny $s$=3};
	
	\draw [very thin,gray,dashed] (.5,0) -- (12,0);
	\draw [very thin,gray,dashed] (.5,2) -- (12,2);
	\draw [very thin,gray,dashed] (.5,4) -- (12,4);
	\draw [very thin,gray,dashed] (.5,6) -- (12,6);

	\foreach \x in {1,3,5,7,9,11}{
		\draw (\x,0) -- (\x+.5,2-.15);
	}
	\foreach \x in {2,4,6,8,10,12}{
	\draw (\x,0) -- (\x-.5,2-.15);
	}

	\foreach \x in {1.5,7.5}{
	\draw (\x,2) -- (\x+2,4-.15);}
	\foreach \x in {3.5,9.5}{
	\draw (\x,2) -- (\x,4-.15);}
	\foreach \x in {5.5,11.5}{
	\draw (\x,2) -- (\x-2,4-.15);}
	\foreach \x in {3.5}{
	\draw (\x,4) -- (\x+3,6-.15);}
	\foreach \x in {9.5}{
	\draw (\x,4) -- (\x-3,6-.15);}

	\foreach \x in {1,2,3,5,6,7}{
	\fill [gray!20] (\x,0) circle [radius=.15];
	\draw[very thin] (\x,0) circle [radius=.15];	
	}
	\foreach \x in {4,8,9,10,11,12}{
	\fill [gray] (\x,0) circle [radius=.15];
	\draw[very thin] (\x,0) circle [radius=.15];
	}
	
	\foreach \x in {1.5,3.5,5.5,7.5,9.5,11.5}{
	\fill [black] (\x,2) circle [radius=.15];	
	}

	\foreach \x in {3.5,9.5}{
	\fill [black] (\x,4) circle [radius=.15];	
	}

	\foreach \x in {6.5}{
	\fill [black] (\x,6) circle [radius=.15];
	}

	\end{tikzpicture}
		\caption{}
		\label{fig:sez_prel:n12_decoding_tree}
	\end{subfigure}
	\hspace{1cm}
	\begin{subfigure}{0.41\columnwidth}
		\begin{tikzpicture}[scale=.4]

\draw [very thin,gray,dashed] (-3.5,1) -- (3.5,1);
\draw [very thin,gray,dashed] (-3.5,-2) -- (3.5,-2);
\draw [very thin,gray,dashed] (-3.5,-5) -- (3.5,-5);

\node at (-4.5,1) {\tiny $s$+1};
\node at (-4.5,-2) {\tiny $s$};
\node at (-4.5,-5) {\tiny $s$-1};

\fill [line width=1mm] (0,1) circle [radius=.2];
\fill [line width=.75mm] (0,-2) circle [radius=.2];
\fill [line width=.5mm] (-2.9,-5) circle [radius=.2];
\fill [line width=.5mm] (0,-5) circle [radius=.2];
\fill [line width=.5mm] (2.9,-5) circle [radius=.2];

\draw [->,very thick] (-.15,0.5) -- (-.15,-1.8) node [left,midway,rotate=0] {$\alpha$};
\draw [->,very thick] (.15,-1.7+.3) -- (.15,0.8) node [right,midway,rotate=0] {$\beta$};

\draw [->,very thick] (-.35-.15,-2.3-.15) -- (-2.85,-4.75) node [left,midway,rotate=0] {$\alpha^{\text{l}}$};
\draw [->,very thick] (-2.6+.15,-4.85+.15) -- (-.1,-2.3) node [right,near start,rotate=0] {};

\draw [<-,very thick] (2.6,-4.85) -- (.1+.3,-2.3-.3) node [left,near start,rotate=0] {$\alpha^{\text{r}}$};
\draw [<-,very thick] (.25,-2.2) -- (2.85-.3,-4.75+.3) node [right,midway,rotate=0] {$\beta^{\text{r}}$};

\draw [<-,very thick] (.1,-2.5) -- (.1,-4.75+.35)  node [left,midway,rotate=0] { };
\draw [<-,very thick] (-.1,-4.75) -- (-.1,-2.65-.15) node [right,midway,rotate=0]  
{$\beta^{\text{c}}$} ;

\node at (-.6,-3.5) {$\alpha^{\text{c}}$};
\node at (-1.5,-4.5) {$\beta^{\text{l}}$};
\end{tikzpicture}
		\caption{}
		\label{fig:sez_prel:tern_message_passing}
	\end{subfigure}
	\caption{(a) Decoding tree for a $\mathcal{P}(12,6)$ polar code and (b) ternary node message passing.}
	\label{fig:sez_prel:n12tree_messagePassing}
\end{figure}

Fig.\ref{fig:sez_prel:n12_decoding_tree} shows the SC decoding tree for the same code, and the message passing criterion in case of ternary nodes is shown in Fig. \ref{fig:sez_prel:tern_message_passing}.
Defining (\ref{eqn:sez_prel:f_bin}) as $f^b$, (\ref{eqn:sez_prel:g_bin}) as $g^b$, and (\ref{eqn:sez_prel:comb_bin}) as $comb^b$, for a ternary node at stage $s$ the decoding rules $\forall i \in \left(0,1,...,\frac{N_s}{3}-1\right)$ are:

\begin{multline}
\alpha^l_i= 2\arctanh \left( \tanh\frac{\alpha_i}{2} \cdot \tanh\frac{\alpha_{i+\frac{N_s}{3}}}{2} \cdot \tanh\frac{\alpha_{i+\frac{2N_s}{3}}}{2}\right)\\
\simeq \varphi \left(\alpha_i\right) \varphi \left(\alpha_{i+\frac{N_s}{3}}\right) \varphi \left(\alpha_{i+\frac{2N_s}{3}}\right) \cdot \\ \cdot \min\left(\left\rvert \alpha_i \right\rvert, \left\rvert \alpha_{i+\frac{N_s}{3}}\right\rvert, \left\rvert \alpha_{i+\frac{2N_s}{3}}\right\rvert\right)
\label{eqn:sez_prel:f_tern} 
\end{multline}
\begin{align}
\alpha^c_i = \left(1-2 \beta^l_i\right)\alpha_i + f^b\left(\alpha_{i+\frac{N_s}{3}},\alpha_{i+\frac{2N_s}{3}}\right)~,
\label{eqn:sez_prel:g1_tern}
\end{align}
\begin{align}
\alpha^r_i = \left(1-2 \beta^l_i\right)\alpha_{i+\frac{N_s}{3}} + \left(1-2 \beta^l_i\oplus\beta^c_i\right)\alpha_{i+\frac{2N_s}{3}}~,
\label{eqn:sez_prel:g2_tern}
\end{align}
\begin{align}
\left[\beta_i, \beta_{i+\frac{N_s}{3}}, \beta_{i+\frac{2N_s}{3}}\right] = \left[ \beta^l_i \oplus \beta^c_i,  \beta^l_i \oplus \beta^r_i, \beta^l_i \oplus \beta^c_i \oplus \beta^r_i \right]~.
\label{eqn:sez_prel:comb_tern}
\end{align}
Similarly to the binary kernel case, we define (\ref{eqn:sez_prel:f_tern}), (\ref{eqn:sez_prel:g1_tern}), (\ref{eqn:sez_prel:g2_tern}) and (\ref{eqn:sez_prel:comb_tern}) as $f^t$, $g_1^t$, $g_2^t$ and $comb^t$ respectively.

%
%
%
%

\section{Multi-kernel codes} \label{sec:codes}

The multi-kernel code construction method proposed in \cite{Gabry_MK} yields substantial error-correction performance gain with respect to puncturing and shortening schemes. Table \ref{table:sez_codes:gain} reports such gain when two multi-kernel codes are compared to codes obtained with the puncturing method in \cite{Niu_puncturing} and the shortening method in \cite{Wang_shortening}, for SC decoding and list SC (SCL) \cite{tal_list} with a list size of $8$. Depending on the target FER, the gain ranges from 0.1 to 1.1 dB.

\begin{figure}[t!]
	\centering
	\includegraphics[width=\columnwidth]{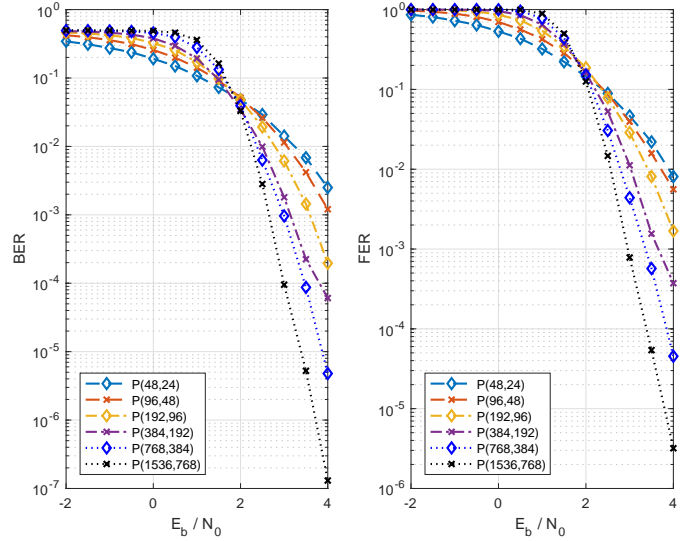}
	\caption{Error-correction performance of binary-ternary mixed polar codes.}
	\label{fig:sez_codes:err_performances}	
\end{figure}

Using the construction method described in \cite{Gabry_MK}, multi-kernel codes have been constructed. 
Their error-correction performance has been simulated through a binary-input additive white Gaussian noise (AWGN) channel with binary phase-shift keying modulation. The bit error rate (BER) and FER curves are shown in Fig. \ref{fig:sez_codes:err_performances}, obtained with SC decoding and LLRs represented in double-precision floating-point format. As discussed in \cite{Gabry_MK,Bioglio_MK}, the Kronecker product is not commutative, and different kernel orders will results in different codes. However, there is currently no theoretical way to identify the best kernel multiplication order: thus, the different kernel orders need to be simulated to identify the one that gives the best error-correction performance. In the remainder of our work, we considered the following codes and kernel orders, obtained with the method described in \cite{Bioglio_MK}:

{\footnotesize
\begin{itemize}
	\item $\mathcal{P}(48,24)$ with $\mathbf{G}=\mathbf{T_3 \otimes T_2 \otimes T_2 \otimes T_2 \otimes T_2}$
	\item $\mathcal{P}(96,48)$ with $\mathbf{G}=\mathbf{T_2 \otimes T_2 \otimes T_2 \otimes T_3 \otimes T_2 \otimes T_2}$
	\item $\mathcal{P}(192,96)$ with $\mathbf{G}=\mathbf{T_3 \otimes T_2 \otimes T_2 \otimes T_2 \otimes T_2 \otimes T_2 \otimes T_2}$
	\item $\mathcal{P}(384,192)$ with $\mathbf{G}=\mathbf{T_3 \otimes T_2\otimes T_2 \otimes T_2 \otimes T_2 \otimes T_2 \otimes T_2 \otimes}$ $\mathbf{\otimes T_2}$
	\item $\mathcal{P}(768,384)$ with $\mathbf{G}=\mathbf{T_2 \otimes T_2 \otimes T_3 \otimes T_2 \otimes T_2 \otimes T_2 \otimes T_2 \otimes}$ $\mathbf{\otimes T_2 \otimes T_2}$
	\item $\mathcal{P}(1536,768)$ with $\mathbf{G}=\mathbf{T_3 \otimes T_2\otimes T_2 \otimes T_2 \otimes T_2  \otimes T_2 \otimes}$ $\mathbf{\otimes  T_2 \otimes T_2 \otimes T_2}$
\end{itemize}
}

\begin{table}[t!]
	\setlength{\extrarowheight}{1.8pt}
	\scriptsize
	\caption{Coding gain for multi-kernel codes with respect to shortening and puncturing schemes.}
	\label{table:sez_codes:gain}
	\begin{center}
		\begin{tabular}{l|cc|cc}
			\toprule
			\multicolumn{5}{c}{$N=72$}\\
			\midrule
			   & \multicolumn{2}{c|}{SC}& \multicolumn{2}{c}{SCL}\\
			FER & $10^{-2}$ & $4\cdot10^{-3}$ & $10^{-2}$ & $2\cdot10^{-3}$ \\
			\cite{Gabry_MK} VS puncturing \cite{Niu_puncturing} & 0.20~dB & 0.45~dB & 0.20~dB& 0.25~dB\\ 
			\cite{Gabry_MK} VS shortening \cite{Wang_shortening} & 0.45~dB  & 0.70~dB & 0.65~dB & 1.10~dB\\
			\midrule
			\multicolumn{5}{c}{$N=48$}\\
			\midrule
			& \multicolumn{2}{c|}{SC}& \multicolumn{2}{c}{SCL}\\
			FER & $10^{-2}$ & $2\cdot10^{-3}$ & $10^{-2}$ & $2\cdot10^{-3}$ \\
			\cite{Gabry_MK} VS puncturing \cite{Niu_puncturing} & 0.10~dB & 0.25~dB & 0.35~dB & 0.50~dB\\
			\cite{Gabry_MK} VS shortening \cite{Wang_shortening} & 0.15~dB & 0.35~dB  &0.75~dB& 0.80~dB\\
			\bottomrule
		\end{tabular}
	\end{center}
\end{table}

\section{Decoder Architecture} \label{sec:HW}
We propose a multi-code semi-parallel SC decoder which supports purely-binary, purely-ternary and binary-ternary mixed construction polar codes. The architecture is sized with a maximum code length $N_{max}$, and can support any code length $N\ge2$ that can be expressed as a combination of binary and ternary kernels, and any code rate. For mixed polar codes, the architecture can decode codes constructed with any kernel order, without knowledge of the code structure at design time.

\begin{figure}[t!]
	\centering
	\includegraphics[width=\columnwidth]{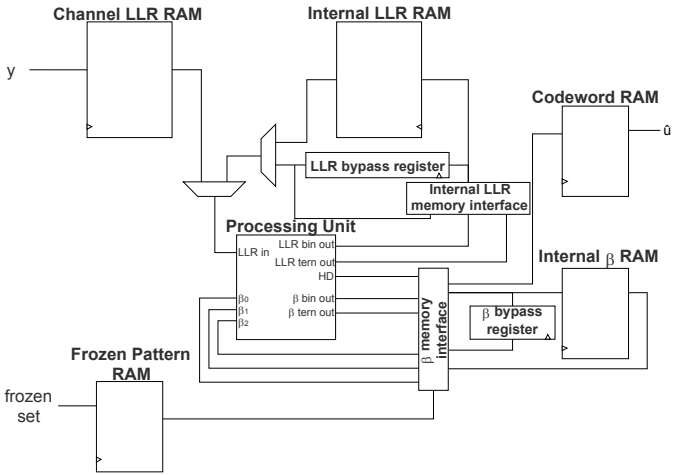}
	\caption{Datapath of the implemented architecture.}
	\label{fig:sez_arch:top_view}	
\end{figure}

The overall decoder architecture is shown in Figure \ref{fig:sez_arch:top_view}. It relies on $P$ processing elements (PEs) implementing (\ref{eqn:sez_prel:f_bin})-(\ref{eqn:sez_prel:comb_tern}), and dedicated memories for channel and internal LLRs, $\beta$ values and candidate codeword. Both channel and internal LLRs are represented on $Q$ bits, $Q_f$ of which are assigned to the fractional part.

Together with the code length, the decoder receives as inputs the following parameters:
\begin{itemize}
 \item information about binary and ternary stages;
 \item memory address offsets for both LLRs and $\beta$ values, relative to the current code length;
 \item number of steps required by each stage to process all inputs given the number $P$ of PEs. This is due to the fact that the decoder has a semi-parallel architecture and, for stages where $N_s>2P$, the number of PEs is not sufficient to elaborate all data in a single clock cycle.
\end{itemize}
In order to simplify and reduce both memory accesses and routing, the architecture has been designed for bit-reversed polar codes \cite{leroux}. This approach allows to dramatically simplify the memory accesses.

\subsection{Data flow}
The channel output $\mathbf{y}$ is initially stored in the \textit{Channel LLR RAM}, while the frozen set $\mathcal{F}$ and the code parameters listed in the previous section are uploaded to their dedicated memories, respectively the \textit{Frozen Pattern RAM} and a set of registers.
For operations involving soft values, the \textit{Processing Unit} receives as input either the channel or the internal LLRs, according to the current stage of the decoding tree. For $comb$ operations (\ref{eqn:sez_prel:comb_bin})-(\ref{eqn:sez_prel:comb_tern}), data read from the \textit{Internal $\beta$ RAM} are used.
Results are stored either in the \textit{Internal LLR RAM} or in the \textit{Internal $\beta$ RAM}, according to the performed operation.
When a leaf node is reached and a hard decision (HD) is performed to decide the value of a bit (\ref{eqn:sez_prel:HD}), the result is stored in the \textit{Codeword RAM}.
The decoding phase ends when the bit associated to the rightmost leaf node is estimated: the decoded codeword $\mathbf{\hat{u}}$ is thus output.

\subsection{Processing Unit} \label{subsec:PU}
The \textit{Processing Unit} (PU) is the computational core of the decoder, where all the operations are performed: $f^b$ (\ref{eqn:sez_prel:f_bin}), $g^b$ (\ref{eqn:sez_prel:g_bin}), $comb^b$ (\ref{eqn:sez_prel:comb_bin}),  $f^t$ (\ref{eqn:sez_prel:f_tern}), $g^t_1$ (\ref{eqn:sez_prel:g1_tern}), $g^t_2$ (\ref{eqn:sez_prel:g2_tern}) and $comb^t$ (\ref{eqn:sez_prel:comb_tern}).
It contains $P$ processing elements (PEs) and $P$ combine blocks (CBs) organized as follows:

\begin{itemize}
	\item $\frac{2}{3}P=P^{b/t}$ binary-ternary mixed PEs, each of them able to compute any $f$ or $g$ operations, both binary and ternary;
	\item $\frac{1}{3}P$ binary PEs, which support only $f^b$ and $g^b$;
	\item $\frac{2}{3}P=P^{b/t}$ binary-ternary mixed CBs which perform both $comb^b$ and $comb^t$;
	\item $\frac{1}{3}P$ binary CBs, which support only $comb^b$.
\end{itemize}

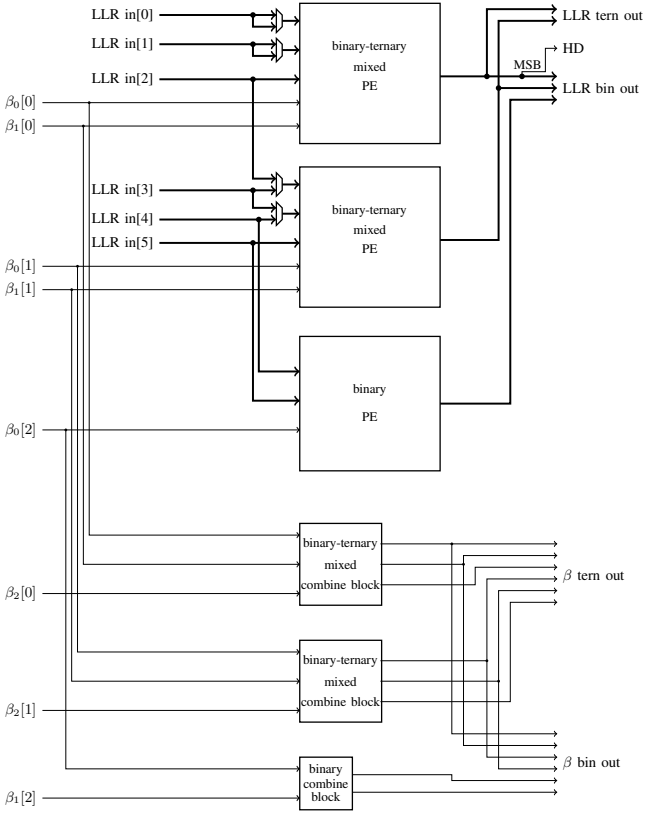
\begin{figure}[t!]
	\resizebox {\columnwidth} {!} {
			\begin{tikzpicture}
	
	\foreach \t in {0,-3.5}{
	\foreach \x in {\t-0,\t-.625}{
	\draw[thick] (0,\x) -- (.125,\x-.125) -- (.125,\x-.375) -- (0,\x-.5) -- cycle;
	\draw[very thick,->] (.125,\x-.25) -- (.5,\x-.25);
	}
	\draw[thick] (.5,\t+.125) rectangle (3+.5,\t-2.875);
	\node at (2,\t-.40625*2) {\footnotesize binary-ternary};
	\node at (2,\t-.40625*3) {\footnotesize mixed};
	\node at (2,\t-.40625*4) {\footnotesize PE};
	}

	\draw[thick] (.5,-7) rectangle (3+.5,-7-2.875);
	\node at (2,-7-.575*2){\footnotesize binary};
	\node at (2,-7-.575*3){\footnotesize PE};
	
	\draw[very thick,->] (3.5,-1.4375) -- (4.5,-1.4375) -- (4.5,-0) -- (6,-0);
	\draw[very thick,->] (3.5,-1.4375-3.5) -- (4.5+.25,-1.4375-3.5) -- (4.5+.25,-.25) -- (6,-.25);
	\node[right] at (6,-.125) {\small LLR tern out};
	
	\draw[very thick,->] (4.5,-1.4375) -- (6,-1.4375);
	\fill (4.5,-1.4375) circle [radius=.06]; 	
	\draw[very thick,->] (4.5+.25,-1.4375-.25) -- (6,-1.4375-.25);
	\fill (4.5+.25,-1.4375-.25) circle [radius=.06]; 
	\draw[very thick,->] (3.5,-1.4375-3.5*2) -- (4.5+.5,-1.4375-3.5*2) --(4.5+.5,-1.4375-.5) -- (6,-1.4375-.5);
	\node[right] at (6,-1.4375-.25) {\small LLR bin out};
	\draw[->] (5.25,-1.4375) --(5.25,-1.4375+.1) --(5.25+.5,-1.4375+.1)-- (5.25+.5,-1.4375+.1+.5)--(6,-1.4375+.1+.5);
 	\node[right] at (6,-1.4375+.1+.5) {\small HD};
 	\node at (5.25+.125,-1.4375+.25) {\footnotesize MSB};
 	\fill (5.25,-1.4375) circle [radius=.06];

 	\node[left] at (-2.5,-.125) {\small LLR in[0]};
 	\draw[very thick,->] (-2.5,-.125) -- (0,-.125);
 	\node[left] at (-2.5,-.75) {\small LLR in[1]};
 	\draw[very thick,->] (-2.5,-.75) -- (0,-.75);
 	\node[left] at (-2.5,-1.5) {\small LLR in[2]};
 	\draw[very thick,->] (-2.5,-1.5) -- (.5,-1.5);
 	\node[left] at (-5,-2) {\small $\beta_0[0]$};
 	\draw[->] (-5,-2) -- (.5,-2);
 	\node[left] at (-5,-2.5) {\small $\beta_1[0]$};
 	\draw[->] (-5,-2.5) -- (.5,-2.5);
 	
	\node[left] at (-2.5,-1-3.5+.625) {\small LLR in[3]};
	\draw[very thick,->] (-2.5,-1-3.5+.625) -- (0,-1-3.5+.625);
	\node[left] at (-2.5,-1-3.5) {\small LLR in[4]};
	\draw[very thick,->] (-2.5,-1-3.5) -- (0,-1-3.5);
	\node[left] at (-2.5,-1.5-3.5) {\small LLR in[5]};
	\draw[very thick,->] (-2.5,-1.5-3.5) -- (.5,-1.5-3.5);
	\node[left] at (-5,-2-3.5) {\small $\beta_0[1]$};
	\draw[->] (-5,-2-3.5) -- (.5,-2-3.5);
	\node[left] at (-5,-2.5-3.5) {\small $\beta_1[1]$};
	\draw[->] (-5,-2.5-3.5) -- (.5,-2.5-3.5);
	\node[left] at (-5,-2-3.5*2) {\small $\beta_0[2]$};
	\draw[->] (-5,-2-3.5*2) -- (.5,-2-3.5*2);
 	
 	\draw[very thick,->] (-.5,-.125) -- (-.5,-.375) -- (0,-.375); 
 	\fill (-.5,-.125) circle [radius=.06]; 	
 	\draw[very thick,->] (-.5,-.125-.625) -- (-.5,-.375-.625) -- (0,-.375-.625); 
 	\fill (-.5,-.125-.625) circle [radius=.06];
 	\draw[very thick,->] (-.5,-1.5) -- (-.5,-3.625)--(0,-3.625);
 	\fill (-.5,-1.5) circle [radius=.06];
 	\draw[very thick,->] (-.5,-3.625-.25) -- (-.5,-3.625-.25-.375) -- (0,-3.625-.25-.375);
 	\fill (-.5,-3.625-.25) circle [radius=.06];
 	
 	\draw[very thick,->] (-.5+.125,-3.625-.25-.625) -- (-.5+.125,-3.625-.25-.875-3)--(.5,-3.625-.25-.875-3);
 	\fill (-.5+.125,-3.625-.25-.625) circle [radius=.06];
 	\draw[very thick,->] (-.5,-3.625-.25-.625-.5) -- (-.5,-3.625-.25-.875-3-.625)--(.5,-3.625-.25-.875-3-.625);
 	\fill (-.5,-3.625-.25-.625-.5) circle [radius=.06];

	\foreach \t in {-11,-13.5}{
	\draw[thick] (.5,\t) rectangle (1.75+.5,\t-1.75);
	\node at (1.375,\t-.4375) {\footnotesize binary-ternary};
	\node at (1.375,\t-.4375*2) {\footnotesize mixed};
	\node at (1.375,\t-.4375*3) {\footnotesize combine block};
	}
	\foreach \t in {-16}{
	\draw[thick] (.5,\t) rectangle (1.125+.5,\t-1.125);
	\node at (1.0625,-16-.28125) {\footnotesize binary};
	\node at (1.0625,-16-.28125*2) {\footnotesize combine};
	\node at (1.0625,-16-.28125*3) {\footnotesize block};
	}
	
	\draw[->] (-4,-2) -- (-4,-11.25) -- (.5,-11.25);
	\fill (-4,-2) circle [radius=.03];
	\draw[->] (-4-.125,-2.5) -- (-4-.125,-11.875) -- (.5,-11.875);
	\fill (-4-.125,-2.5) circle [radius=.03];
	
	\draw[->] (-4-.25,-2-3.5) -- (-4-.25,-11.25-2.5) -- (.5,-11.25-2.5);
	\fill (-4-.25,-2-3.5) circle [radius=.03];
	\draw[->] (-4-.125-.25,-2.5-3.5) -- (-4-.25-.125,-11.875-2.5) -- (.5,-11.875-2.5);
	\fill (-4-.125-.25,-2.5-3.5) circle [radius=.03];

	\draw[->] (-4-.25-.25,-2-3.5*2) -- (-4-.25-.25,-11.25-2.5*2) -- (.5,-11.25-2.5*2);
	\fill (-4-.25-.25,-2-3.5*2) circle [radius=.03];

	\draw[->] (-5,-12.5) -- (.5,-12.5);
	\node[left] at (-5,-12.5) {\small $\beta_2[0]$};
	\draw[->] (-5,-15) -- (.5,-15);
	\node[left] at (-5,-15) {\small $\beta_2[1]$};
	
	\draw[->] (-5,-16.875) -- (.5,-16.875);
	\node[left] at (-5,-16.875) {\small $\beta_1[2]$};
	
	\draw[->] (1.725+.5,-11.4375) -- (6,-11.4375);
	\draw[->] (1.725+.5,-11.4375-0.4375) -- (4,-11.4375-0.4375)-- (4,-11.4375-.25)--(6,-11.4375-.25);
	\draw[->] (1.725+.5,-11.4375-0.4375*2) -- (4.25,-11.4375-0.4375*2)-- (4.25,-11.4375-.25*2)--(6,-11.4375-.25*2);
	
	\draw[->] (1.725+.5,-11.4375-2.5) -- (4.5,-11.4375-2.5)-- (4.5,-11.4375-.25*3)--(6,-11.4375-.25*3);
	\draw[->] (1.725+.5,-11.4375-0.4375-2.5) -- (4.75,-11.4375-0.4375-2.5)-- (4.75,-11.4375-.25*4)--(6,-11.4375-.25*4);
	\draw[->] (1.725+.5,-11.4375-0.4375*2-2.5) -- (5,-11.4375-0.4375*2-2.5)--(5,-11.4375-.25*5)--(6,-11.4375-.25*5);
	
	\draw[->] (1.125+.5,-16.75+.375) -- (3.75,-16.75+.375)--(3.75,-16.75+.25)--(6,-16.75+.25);
	\draw[->] (1.125+.5,-16.75) -- (6,-16.75);
	
	\draw[->] (3.75,-11.4375)--(3.75,-16.75+.25*5)--(6,-16.75+.25*5);
	\fill (3.75,-11.4375) circle [radius=.03];
	\draw[->] (4,-11.4375-0.4375) -- (3.75+.25,-16.75+.25*4)-- (6,-16.75+.25*4);
	\fill (4,-11.4375-0.4375) circle [radius=.03];
	\draw[->] (4.5,-11.4375-2.5) -- (3.75+.25*3,-16.75+.25*3)-- (6,-16.75+.25*3);
	\fill (4.5,-11.4375-2.5) circle [radius=.03];
	\draw[->] (4.75,-11.4375-0.4375-2.5) -- (3.75+.25*4,-16.75+.25*2)-- (6,-16.75+.25*2);
	\fill (4.75,-11.4375-0.4375-2.5) circle [radius=.03];
	
	\node[right] at (6,-12-.125) {\small $\beta$ tern out};
	\node[right] at (6,-16-.125) {\small $\beta$ bin out};

	\end{tikzpicture}
	}
	\caption{Example of \textit{Processing Unit} with $P=3$.}
	\label{fig:sez_arch:PU}
\end{figure}

Since it has been observed that between binary and ternary operations there are common computations, mixed PEs are used to increase resource sharing, at the cost of a multiplexing operation; additional purely-binary PEs are used to align the number of used inputs both for binary and ternary operations.
Thus the maximum number of elaborated soft inputs is fixed to $2P=3P^{b/t}$, while the results are either $P$ or $P^{b/t}$ LLRs: in fact it can be noticed that the number of operations simultaneously performed is $P$ in the binary case and $P^{b/t}$ in the ternary one. For binary operations each $i$-th PE elaborates the $2i$-th and $(2i+1)$-th LLR inputs, while for ternary ones each $i$-th mixed PE uses LLRs corresponding to indices $3i$, $3i+1$ and $3i+2$.
The same holds for CBs.
From the last considerations $P$ must be a multiple of $3$; an example of PU with $P=3$ is shown in Fig. \ref{fig:sez_arch:PU}.

Although there are situations in which not all PEs are performing useful computations, $2P$ inputs are nevertheless elaborated and stored in the corresponding memory. Unnecessary data are subsequently ignored in the final estimation: this happens for stages $s$ where $N_s$ is not a multiple of $2P$. 
The impact of two different LLR representations on the implementation cost of the PU has been evaluated: we have in fact designed PEs with both 2's complement and sign and magnitude representations. FPGA synthesis results have shown that the sign and magnitude binary PE has $14\%$ lower resource requirements and $20\%$ shorter critical path than the 2's complement one, while the sign and magnitude mixed PE has similar resource requirements and $23\%$ shorter critical path than the 2's complement one. Thus, all LLRs in the proposed decoder are represented with sign and magnitude.


\subsubsection{Binary Processing Elements}
The architecture of binary PEs is the one proposed in \cite{leroux}. Let us call $\alpha_a$ and $\alpha_b$ the input LLRs. 
For the hardware-friendly version of $f^b$ (\ref{eqn:sez_prel:f_bin}) operation the result computation is straightforward: 

\begin{equation}
\varphi(\alpha^b_f)=\varphi(\alpha_a)\oplus \varphi(\alpha_b)~,
\label{eqn:sez_PEb:sign_f}
\end{equation}
\begin{equation}
|\alpha^b_f|=\min(|\alpha_a|,|\alpha_b|)~ ,
\label{eqn:sez_PEb:magnitude_f}
\end{equation}
where $\alpha^b_f$ is the $f^b$ operation result.
Analyzing the complete truth table both for sign $\varphi(\alpha^b_g)$ and magnitude $|\alpha^b_g|$ of $g^b$ (\ref{eqn:sez_prel:g_bin}), its resulting equations are:

\begin{equation}
\varphi(\alpha^b_g)=\overline{\gamma_{ab}} \cdot \varphi(\alpha_b)\, +\, \gamma_{ab}\cdot(u_0 \oplus \varphi(\alpha_a))~,
\label{eqn:sez_PEb:sign_g}
\end{equation}
\begin{equation}
|\alpha^b_g|=\max(|\alpha_a|,|\alpha_b|) \, + \, (-1)^{\chi}\min(|\alpha_a|,|\alpha_b|)~,
\label{eqn:sez_PEb:mag_g}
\end{equation}

where

\begin{equation}
\gamma_{ab} =
\begin{cases}
1 & 
\mbox{if $|\alpha_a|>|\alpha_b|$~,}\\
0 &  \mbox{otherwise~,}
\end{cases}
\label{eqn:sez_PEb:g_gamma_ab}
\end{equation}	
\begin{equation}
\chi=u_0 \oplus \varphi(\alpha_a) \oplus \varphi(\alpha_b)~.
\label{eqn:sez_PEb:g_chi}
\end{equation}

\begin{figure}[t!]
	\includegraphics[width=\columnwidth]{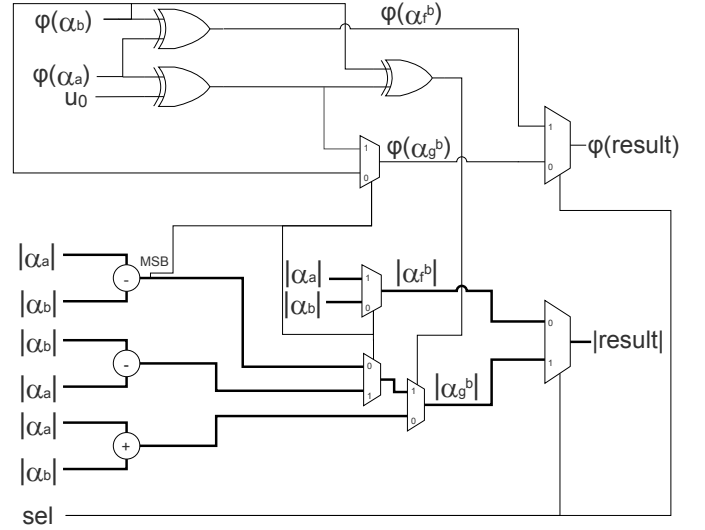}
	\caption{Datapath of a binary PE.}
	\label{fig:sez_PEb:datapath}
\end{figure}

This architecture is shown in Figure \ref{fig:sez_PEb:datapath}\,. Adders and subtractors saturate their result if outside the available range.

\subsubsection{Binary-ternary mixed Processing Elements} \label{sec:BTPE}

An analysis analogous to the binary case has been conducted on $f^t$ (\ref{eqn:sez_prel:f_tern}), $g^t_1$ (\ref{eqn:sez_prel:g1_tern}) and $g^t_2$ (\ref{eqn:sez_prel:g2_tern}). The resulting equations are the following:

\begin{equation}
\varphi(\alpha^t_f)=\varphi(\alpha_a)\oplus \varphi(\alpha_b) \oplus \varphi(\alpha_c)~,
\label{eqn:sez_PEbt:sign_f}
\end{equation}
\begin{equation}
|\alpha^t_f|=\min(|\alpha_a|,|\alpha_b|,|\alpha_c|)~,
\label{eqn:sez_PEbt:magnitude_f}
\end{equation}

\begin{equation}
\varphi(\alpha_{g_1}^t)=\overline{\gamma_{g_1}} \cdot (\varphi(\alpha_b)\oplus \varphi(\alpha_c))\, +\, \gamma_{g_1}\cdot(u_0 \oplus \varphi(\alpha_a))~,
\label{eqn:sez_PEbt:sign_g1}
\end{equation}
\begin{multline}
|\alpha_{g_1}^t|=\max(|\alpha_a|,\min(|\alpha_b|,|\alpha_c|)) \, +
\\
+ \, (-1)^{\chi_{g1}}\min(|\alpha_a|,\min(|\alpha_b|,\alpha_c|))~,
\label{eqn:sez_PEbt:mag_g1}
\end{multline}
\begin{equation}
\varphi(\alpha_{g_2}^t)=\overline{\gamma_{g_2}} \cdot (u_0 \oplus u_1 \oplus \varphi(\alpha_c))\, +\, \gamma_{g_2}\cdot(u_0 \oplus \varphi(\alpha_b))~,
\label{eqn:sez_PEbt:sign_g2}
\end{equation}
\begin{equation}
|\alpha_{g_2}^t|=\max(|\alpha_b|,|\alpha_c|) \, + \, (-1)^{\chi_{g2}}\min(|\alpha_b|,|\alpha_c|)~,
\label{eqn:sez_PEbt:mag_g2}
\end{equation}

where

\begin{equation}
\gamma_{g_1} =
\begin{cases}
1 & 
\mbox{if $|\alpha_a|>\min(|\alpha_b|,|\alpha_c|)$}\\
0 &  \mbox{otherwise}
\end{cases}~,
\label{eqn:sez_PEbt:g1_gamma}
\end{equation}	
\begin{equation}
\chi_{g1}=u_0 \oplus \varphi(\alpha_a) \oplus \varphi(\alpha_b) \oplus \varphi(\alpha_c)~,
\label{eqn:sez_PEbt:g1_chi}
\end{equation}

\begin{equation}
\gamma_{g_2} =
\begin{cases}
1 & 
\mbox{if $|\alpha_b|>|\alpha_c|$}\\
0 &  \mbox{otherwise}
\end{cases}~,
\label{eqn:hw:tern:g2_gamma}
\end{equation}
\begin{equation}
\chi_{g2}=u_1 \oplus \varphi(\alpha_b) \oplus \varphi(\alpha_c)~.
\label{eqn:sez_PEbt:g2_chi}
\end{equation}

\begin{figure}[t!]
	\includegraphics[width=\columnwidth]{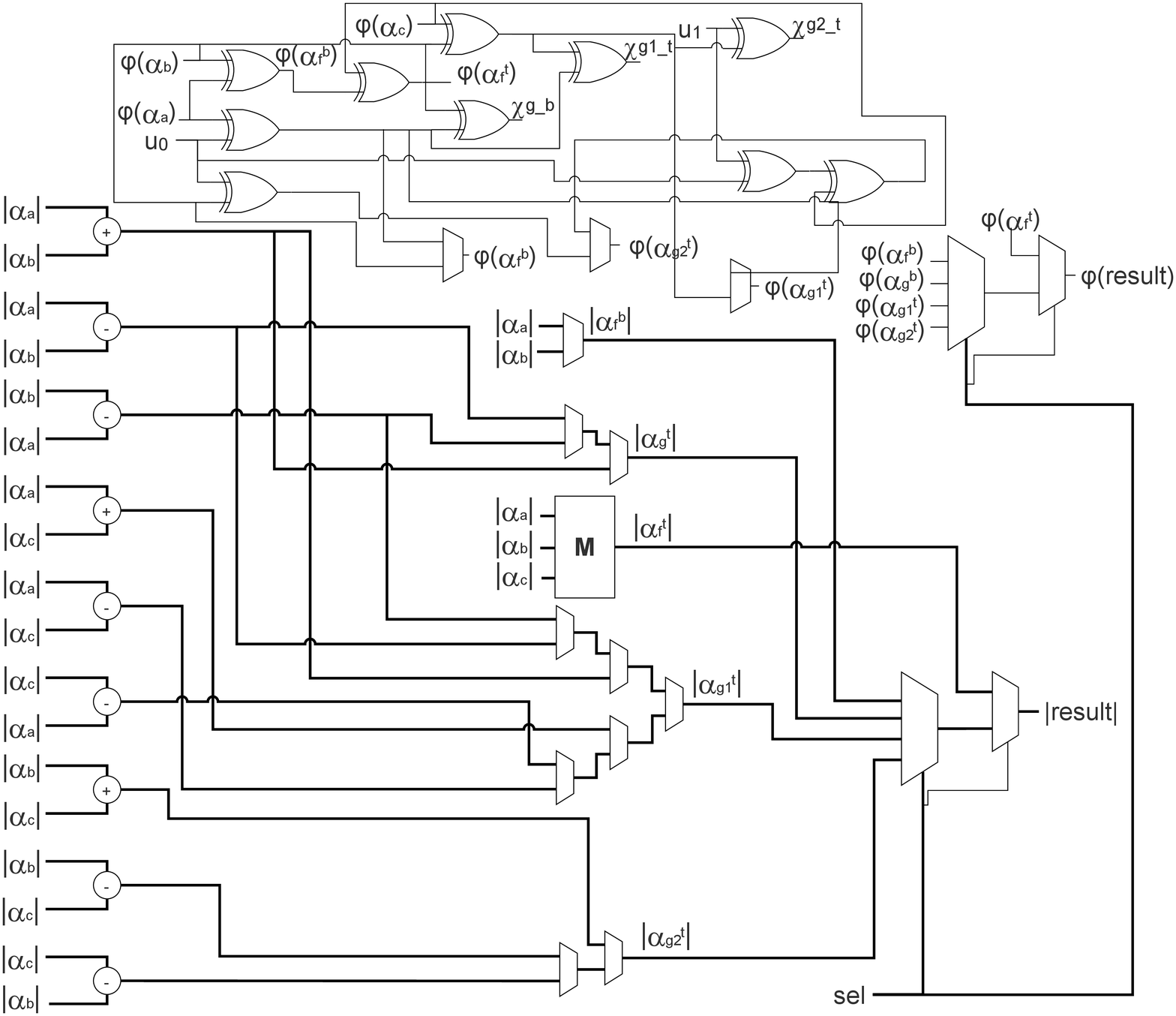}
	\caption{Datapath of a binary-ternary mixed PE.}
	\label{fig:sez_PEbt:datapath}
\end{figure}

The circuit implementing these operations is shown in Figure \ref{fig:sez_PEbt:datapath}, where again adders and subtractors can saturate the result. The $M$ block is a combination of pruned multiplexers selecting the minimum absolute value according to the already computed selection signals, which correspond to the most significant bits of the output of the subtractors. 

Mixed PEs perform both binary and ternary operations, and need to select their input accordingly. Thus, LLR multiplexing logic is inserted at their input. This logic consists of two Q-bit multiplexers for each mixed PE. 

%

\subsubsection{Combine blocks}
Both binary and binary-ternary mixed CBs are composed of XOR gates implementing $comb^b$ and $(comb^b)\overline{sel} + (comb^t)sel$ respectively, where $sel$ is the binary/ternary selector.

\subsection{Memory system} \label{subsec:memsys}
While efficient in terms of resource usage, register-based approaches like \cite{hashemi_SSCL_TCASI} lead to excessive area occupation. Thus, this design foresees the usage of SRAM banks. The width of these memories is different from that of memories in a purely-binary decoder design, since they have to accommodate ternary operations and their concurrent input and output volume. Additionally, for \textit{Internal LLR RAM} a three-bank solution has been implemented, since ternary-kernel functions are supported: for purely-binary decoders two banks would have been sufficient.

\subsubsection{Channel LLR RAM}
This memory stores the LLRs coming from the channel. Each memory word is $2P\cdot Q$ long, since for each operation involving LLRs $2P$ of them are required by the PU. Its depth is $D_{LLR\, ch}=\left\lceil \frac{N_{max}}{2P} \right\rceil $. This memory uses two separate ports, one for reading and one for writing.

\subsubsection{Internal LLR RAM}
It contains the partial results of $f$ and $g$ operations. Similarly to the \textit{Channel LLR RAM}, the parallelism must be $2P\cdot Q$. The computation of the depth $D_{LLR\, int}$ takes into account that for each decoding stage only one LLR vector must be stored: once the node which took as input the computed LLR has generated its output $\beta$, that soft value will be no longer used and can be overwritten. 
In addition, for stage $s=0$ it is not needed to memorize the result since the hard decision is performed in the same clock cycle.

The memory depth is computed as:

$$\displaystyle D_{LLR\, int}=\sum_{s=1}^{\log_2( N_{max})-1}\left\lceil \frac{N_{max}}{2^s \cdot 2P} \right\rceil ~.$$
Also for this memory two separate ports for reading and writing are required.

It is possible to rearrange the \textit{Internal LLR RAM} with a bank structure. However, due to the variable number of data that needs to be written, depending on the stage being binary or ternary, four banks with two different widths should be implemented. This would incur significant control and addressing overhead, with no tangible advantage with respect to the proposed structure.
More details on the handling of different result sizes are given in Section \ref{subsection:mem_int}.


\subsubsection{Internal $\beta$ RAM}
This memory stores all $\beta$ values computed inside the decoding tree; it is organized in three banks, which share the same input writing bus:
\begin{itemize}
	\item \textsc{bank0} for $\beta_0$: it is equal to $\beta^l$ in both binary and ternary cases;
	\item \textsc{bank1} for $\beta_1$: it is equal to $\beta^r$ for binary stages, while for ternary ones it represents $\beta^c$;
	\item \textsc{bank2} for $\beta_2$: it corresponds to the ternary stages $\beta^r$.
\end{itemize}

The bank organization is fundamental for parallel data reading in $g^t_2$, $comb^b$ and $comb^t$ operations. Each bank has a width of $2P$ since results of $comb$ operations are on $2P$ bits, while their depths $D_{\beta\, int}$ are equal to: 
$$ \displaystyle D_{\beta\, int}=\sum_{s=0}^{\log_2( N_{max})-1}\left\lceil \frac{N_{max}}{2^s \cdot 2P} \right\rceil~.$$

\subsubsection{Codeword RAM}
It is used to store the decoder output $\mathbf{\hat{u}}$, composed by the HDs performed at the leaf nodes. Its width $W_{cod}$ is a design choice independent from all other parameters, while the depth is
\begin{equation*}
D_{cod}=\left\lceil \frac{N_{max}}{W_{cod}} \right\rceil~.
\end{equation*}
\subsubsection{Frozen Pattern RAM}
It stores the frozen set, where each of $N_{max}$ bits identifies if the corresponding bit-channel is frozen or not. The memory width $W_{frozen}$ is an independent design choice, while the depth can be expressed as $$D_{frozen}=\left\lceil \frac{N_{max}}{W_{frozen}} \right\rceil~.$$

Table \ref{table:sez_memys:mem_sizes} reports the breakdown of the memory requirements for the proposed decoder with various $N_{max}$, $P$ and $Q$ combinations. To correctly evaluate the memory overhead brought by the multi-kernel approach, the memory sizes for purely binary polar decoders with similar parameters have been detailed as well. It can be seen that most of the additional memory bits can be found in the internal $\beta$ memory.
\begin{table}[t!]
	{\scriptsize
	\setlength{\extrarowheight}{1.8pt}
	\caption{Memory requirements for various decoder parameters, considering both a multi-kernel (MK) and a purely binary (PB) approach.}
	\label{table:sez_memys:mem_sizes}
\begin{center}
	\begin{tabular}{c|cc|cc|cc}
		\toprule
		& MK & PB & MK & PB & MK & PB\\
		\midrule
		$N_{max}$&$4096$&$4096$&$1024$&$1024$&$256$&$256$\\
		$P$&$120$&$128$&$60$&$64$&$18$&$16$\\
		$Q$&$7$&$7$&$6$&$6$&$5$&$5$\\
		\midrule
		& [bit] &  [bit] &  [bit] &  [bit] &  [bit] &  [bit]  \\
		Channel LLR RAM &30240&28672&6480&6144&1440&1280\\
		Internal LLR RAM  &43680&39424&11520&9984&1980&1760\\
		Internal $\beta$ RAM  &31680&19456&9000&5376&2052&1216\\
		Codeword RAM  &4096&4096&1024&1024&256&256\\
		Frozen Pattern RAM  &4096&4096&1024&1024&256&256\\
		&&&&&&\\
		Total  &113792&95744&29048&23552&5984&4768\\
		\bottomrule
	\end{tabular}
\end{center}}
\end{table}

\subsection{Memory interfaces}
\label{subsection:mem_int}
Two interfacing modules are required to adapt the inherent parallelism of the memories to that of the PU.

\subsubsection{Internal LLR memory interface}
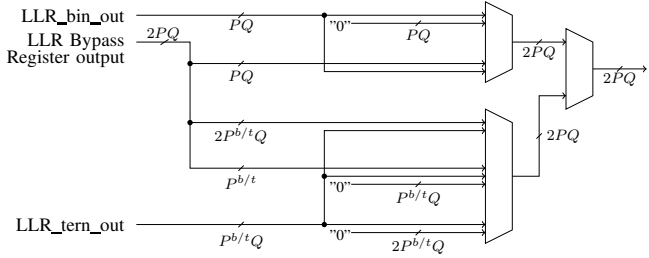
\begin{figure}[t!]
	\resizebox {\columnwidth} {!} {
			\begin{tikzpicture}
	\node[left] at (-2.6,-2) {LLR\_bin\_out};
	\draw [->] (-2.5,-2) -- (4,-2);
	\fill (1,-2) circle [radius=.05];
	\draw [->] (1,-2) -- (1,-3.05) -- (4,-3.05); 
	
	\node[left] at (-2.6,-2.5) {LLR Bypass};
	\node[left] at (-2.6,-2.8) {Register output};
	\draw [->] (-2.5,-2.5) -- (-1.5,-2.5) -- (-1.5,-2.9) -- (4,-2.9);
	\fill (-1.5,-2.9) circle [radius=.05];
	\draw [->] (-1.5,-2.9) -- (-1.5,-4) -- (4,-4);
	\fill (-1.5,-4) circle [radius=.05];
	\draw [->] (-1.5,-4) -- (-1.5,-4.85) -- (4,-4.85);
	
	\node at (1.3,-2.2) {\footnotesize "0"};
	\draw [->] (1.5,-2.15) -- (4,-2.15);
	
	\draw (4,-1.75) -- (4,-3.25) -- (4.5,-3) -- (4.5,-2) -- cycle; 
	
	\draw (4,-3.75) -- (4,-6.25) -- (4.5,-6) -- (4.5,-4) -- cycle;
	
	\draw [->] (4.5,-2.5) -- (5.5,-2.5);
	\draw (5-.05,-2.5-.05)--(5+.05,-2.5+.05);
	\node at (5,-2.5-.2) {\footnotesize $2PQ$};
	\draw [->] (4.5,-5) -- (5,-5) -- (5,-3.5) -- (5.5,-3.5);
	\draw (5-.05,-4.25-.05) -- (5+.05,-4.25+.05);
	\node at (5+.45,-4.25) {\footnotesize $2PQ$};
	\draw [->] (6,-3) -- (7,-3);
	\draw (6.45,-3.05) -- (6.55,-2.95);
	\node at (6.5,-3.2) {\footnotesize $2PQ$};
	
	\draw (5.5,-2.25) -- (5.5,-3.75) -- (6,-3.5) -- (6,-2.5) -- cycle;

	\node[left] at (-2.6,-5.9) {LLR\_tern\_out};
	\draw [->] (-2.5,-5.9) -- (4,-5.9);
	\node at (1.3,-6.1) {\footnotesize "0"};
	\draw [->] (1.5,-6.05) -- (4,-6.05);
	\node at (1.3,-5.2) {\footnotesize "0"};
	\draw [->] (1.5,-5.15) -- (4,-5.15);

	\fill (1,-5.9) circle [radius=.05];
	\draw [->] (1,-5.9) -- (1,-5) -- (4,-5); 
	\fill (1,-5) circle [radius=.05];
	\draw [->] (1,-5) -- (1,-4.15) -- (4,-4.15); 
	
	\draw (2.7,-2.2) -- (2.8,-2.1);
	\node at (2.75,-2.4) {\footnotesize $P Q$};
	\draw (2.7,-5.2) -- (2.8,-5.1);
	\node at (2.75,-5.4) {\footnotesize $P^{b/t} Q$};
	\draw (2.7,-6.1) -- (2.8,-6);
	\node at (2.75,-6.3) {\footnotesize $2P^{b/t} Q$};
	
	\draw (-.6,-2-.05) -- (-.5,-2+.05);
	\node at (-.55,-2.2) {\footnotesize $P Q$};
	\draw (-.6,-5.9-.05) -- (-.5,-5.9+.05);
	\node at (-.55,-5.9-.3) {\footnotesize $P^{b/t} Q$};
	\draw (-.6,-2.9-.05) -- (-.5,-2.9+.05);
	\node at (-.55,-2.75-.4) {\footnotesize $PQ$};
	\draw (-.6,-4-.05) -- (-.5,-4+.05);
	\node at (-.555,-3.85-.4) {\footnotesize $2P^{b/t}Q$};
	\draw (-.6,-4.85-.05) -- (-.5,-4.85+.05);
	\node at (-.55,-4.7-.4) {\footnotesize $P^{b/t}$};
	
	\draw (-2.05,-2.5-.05) -- (-1.95+.05,-2.5+.05);
	\node at (-2,-2.35) {\footnotesize $2PQ$};
	
	\end{tikzpicture}
	}
	\caption{\textit{Internal LLR RAM} interface circuit.}
	\label{fig:sez_mem_int:llr_interface}	
\end{figure}

 Fig. \ref{fig:sez_mem_int:llr_interface} shows the interface circuit.
It is tasked with choosing, during write operations, which part of the memorized word has to be overwritten. In fact, the results of $f$ and $g$ operations are $P$ or $P^{b/t}$ LLR, for binary and ternary cases respectively, while the width of the LLR memories is $2P=3P^{b/t}$. Each memory location takes two or three clock cycles to be overwritten with useful data. So, at tree stages where $N_s>2P$ and the PU takes more than one clock cycle to process them, the following steps are performed:
\begin{itemize}
	\item For binary stages:
	\begin{enumerate}
		\item The $2i$-th operation result ($P$ LLRs) is stored in the memory together with $QP^b$ appended zeros;
		\item The $(2i+1)$-th operation result is stored after the $P$ most significant bits of the previously written word, so that the padding zeros are overwritten and the new stored word contains the $P$ results of both the $2i$-th and $(2i+1)$-th operations.
	\end{enumerate}
	\item For ternary stages:
	\begin{enumerate}
		\item The $3i$-th operation result ($P^{b/t}$ LLRs) is stored in the memory together with $2QP^{b/t}$ appended zeros.
		\item The $(3i+1)$-th operation result is stored after the $QP^{b/t}$ most significant bits of the previously written word. The new word contains the $P^{b/t}$ results of both the $3i$-th and $(3i+1)$-th operations;
		\item The $(3i+2)$-th operation result is stored after the previously written $2QP^{b/t}$ bits, completing the $3P^{b/t}=2P$ LLR word.
	\end{enumerate}
\end{itemize} 
To overwrite only parts of the previously written word, the bypass buffer output is used.
When $N_s\le2P$, the results are stored in the first part of the word as usual; the remaining bits are not considered in subsequent operations.

\subsubsection{$\beta$ memory interface}
\begin{figure}[t!]
	\resizebox {\columnwidth} {!} {
			\begin{circuitikz}
	\draw (-1,2) -- (-1,1) -- (12.5,1) -- (12.5,2);
	\node at (5.75,1.85) {\Large Internal $\beta $ RAM};
	
	\node[right] at (-3.5,.2) {\Large $\beta$ buffer out};
	\draw [->] (-3,0) -- (0,0) -- (0,-.5);
	\draw [->] (-3,0) -- (4,0) -- (4,-.5);
	\fill (0,0) circle [radius=.05];
	\draw [->] (4,0) -- (8,0) -- (8,-.5);
	\fill (4,0) circle [radius=.05];
	\draw (-.5-.1,-.1) -- (-.5+.1,.1);
	\node at (-.5,.2) {\normalsize $2P$};

	\draw (-.25,-.5) -- (.75,-.5) -- (.5,-.75) -- (0,-.75) -- cycle;
	
	\node at (.5,1.25) {\Large \texttt{r\_data\_0}};
	\draw [->] (.5,1) -- (.5,-.5);
	\draw (.5-.1,0.25-.1)--(.5+.1,0.25+.1);
	\node[right] at (.5,0.25) {\normalsize $2P$};
	\draw [->] (.25,-.75) -- (.25,-1.5) -- (.25+1+1.5,-1.5) -- (.25+1+1.5,-4.5);
	\fill (.25,-1.5) circle [radius=.05];
	\draw [->] (.25,-1.5) -- (.25,-4.5);
	\fill (.25+.5,-1.5) circle [radius=.05];
	\draw [->] (.2+.55,-1.5) -- (.25+.5,-4.5);
	
	\fill (.25+1.5,-1.5) circle [radius=.05];
	\draw [->] (.25+1.5,-1.5) -- (.25+1.5,-4.5);
	\fill (.25+.5+1.5,-1.5) circle [radius=.05];
	\draw [->] (.2+.55+1.5,-1.5) -- (.25+.5+1.5,-4.5);
	
	\draw (0,-4.5) -- (1,-4.5) -- (.75,-4.75) -- (.25,-4.75) -- cycle;
	\draw [->] (0.5,-4.75) -- (0.5,-5.25) --(1.375-.25,-5.25)-- (1.375-.25,-6); 
	\draw (.5-.1,-5-.1) -- (.5+.1,-5+.1);
	\node[right] at (.5,-5) {\normalsize $P$};
	
	\draw (1.5,-4.5) -- (3,-4.5) -- (2.75,-4.75) -- (1.75,-4.75) -- cycle;
	\draw [->] (2.25,-4.75) -- (2.25,-5.25) -- (1.375+.125,-5.25) --(1.375+.125,-6); 
	\draw (2.25-.1,-5-.1) -- (2.25+.1,-5+.1);
	\node[right] at (2.25,-5) {\normalsize $P^{b/t}$};
	\draw [->] (1.375+.25,-5.6)--(1.375+.25,-6);
	\node[right] at (1.375,-5.5) {\normalsize "0"};
	\draw (1.375+.25-.1,-5.75-.1) -- (1.375+.25+.1,-5.75+.1);
	\node[right] at (1.375+.25,-5.75) {\normalsize $P-P^{b/t}$};
	
	\draw (0.875,-6) -- (1.875,-6) -- (1.625,-6.25) -- (1.125,-6.25) -- cycle;
	
	\draw [->] (1.375,-6.25) -- (1.375,-7);
	\node at (1.375,-7.3) {\Large $\beta_{0}$};
	\draw (1.375-.1,-6.5-.1) -- (1.375+.1,-6.5+.1);
	\node[right] at (1.375,-6.5) {\normalsize $P$};
	
	\draw (-.25+4,-.5) -- (.75+4,-.5) -- (.5+4,-.75) -- (0+4,-.75) -- cycle;
	
	\node at (.5+4,1.25) {\Large \texttt{r\_data\_1}};
	\draw [->] (.5+4,1) -- (.5+4,-.5);
	\draw (.5-.1+4,0.25-.1)--(.5+.1+4,0.25+.1);
	\node[right] at (.5+4,0.25) {\normalsize $2P$};
	\draw [->] (.25+4,-.75) -- (.25+4,-1.5) -- (.25+1+1.5+4,-1.5) -- (.25+1+1.5+4,-4.5);
	\fill (.25+4,-1.5) circle [radius=.05];
	\draw [->] (.25+4,-1.5) -- (.25+4,-4.5);
	\fill (.25+.5+4,-1.5) circle [radius=.05];
	\draw [->] (.2+.55+4,-1.5) -- (.25+.5+4,-4.5);
	
	\fill (.25+1.5+4,-1.5) circle [radius=.05];
	\draw [->] (.25+1.5+4,-1.5) -- (.25+1.5+4,-4.5);
	\fill (.25+.5+1.5+4,-1.5) circle [radius=.05];
	\draw [->] (.2+.55+1.5+4,-1.5) -- (.25+.5+1.5+4,-4.5);
	
	\draw (0+4,-4.5) -- (1+4,-4.5) -- (.75+4,-4.75) -- (.25+4,-4.75) -- cycle;
	\draw [->] (0.5+4,-4.75) -- (0.5+4,-5.25) --(1.375-.25+4,-5.25)-- (1.375-.25+4,-6); 
	\draw (.5-.1+4,-5-.1) -- (.5+.1+4,-5+.1);
	\node[right] at (.5+4,-5) {\normalsize $P$};
	
	\draw (1.5+4,-4.5) -- (3+4,-4.5) -- (2.75+4,-4.75) -- (1.75+4,-4.75) -- cycle;
	\draw [->] (2.25+4,-4.75) -- (2.25+4,-5.25) -- (1.375+.125+4,-5.25) --(1.375+.125+4,-6); 
	\draw (2.25-.1+4,-5-.1) -- (2.25+.1+4,-5+.1);
	\node[right] at (2.25+4,-5) {\normalsize $P^{b/t}$};
	\draw [->] (1.375+.25+4,-5.6)--(1.375+.25+4,-6);
	\node[right] at (1.375+4,-5.5) {\normalsize "0"};
	\draw (1.375+.25-.1+4,-5.75-.1) -- (1.375+.25+.1+4,-5.75+.1);
	\node[right] at (1.375+.25+4,-5.75) {\normalsize $P-P^{b/t}$};
	
	\draw (0.875+4,-6) -- (1.875+4,-6) -- (1.625+4,-6.25) -- (1.125+4,-6.25) -- cycle;
	
	\draw [->] (1.375+4,-6.25) -- (1.375+4,-7);
	\node at (1.375+4,-7.3) {\Large $\beta_{1}$};
	\draw (1.375-.1+4,-6.5-.1) -- (1.375+.1+4,-6.5+.1);
	\node[right] at (1.375+4,-6.5) {\normalsize $P$};

	\draw (-.25+8,-.5) -- (.75+8,-.5) -- (.5+8,-.75) -- (0+8,-.75) -- cycle;
	
	\node at (.5+8,1.25) {\Large \texttt{r\_data\_3}};
	\draw [->] (.5+8,1) -- (.5+8,-.5);
	\draw (.5-.1+8,0.25-.1)--(.5+.1+8,0.25+.1);
	\node[right] at (.5+8,0.25) {\normalsize $2P$};
	\draw [->] (.25+8,-.75) -- (.25+8,-1.5) -- (.25+1+8,-1.5) -- (.25+1+8,-4.5);
	
	\fill (.25+8,-1.5) circle [radius=.05];
	\draw [->] (.25+8,-1.5) -- (.25+8,-4.5);
	\fill (.25+.5+8,-1.5) circle [radius=.05];
	\draw [->] (.2+.55+8,-1.5) -- (.25+.5+8,-4.5);

	\draw (1.5+8-1.5,-4.5) -- (3+8-1.5,-4.5) -- (2.75+8-1.5,-4.75) -- (1.75+8-1.5,-4.75) -- cycle;
	\draw [->] (2.25+8-1.5,-4.75)--(2.25+8-1.5,-7);
	
	\node at (2.25+8-1.5,-7.3) {\Large $\beta_{2}$};
	\draw (2.25+8-1.5-.1,-6.5-.1) -- (2.25+8-1.5+.1,-6.5+.1);
	\node[right] at (2.25+8-1.5,-6.5) {\normalsize $P^{b/t}$};


	\node at (10.5,-7.3) {\Large $\beta^{bin}_{out}$};
	\node at (11.5,-7.3) {\Large $\beta^{tern}_{out}$};
	\node at (12.5,-7.3) {\Large HD};
	\node[right] at (13,-6.5+.2) {\Large frozen bit};
	
	\draw (10.5,-4.5) -- (11.5,-4.5) -- (11.75,-4.75) -- (10.25,-4.75) --cycle;
	
	\draw [->] (10.5,-7) -- (10.5,-4.75);
	\draw [->] (11.5,-7) -- (11.5,-4.75);

	\draw (10.5-.1,-6.5-.1) -- (10.5+.1,-6.5+.1);
	\node[right] at (10.5,-6.5) {\normalsize $2P$};
	\draw (11.5-.1,-6.5-.1) -- (11.5+.1,-6.5+.1);
	\node[right] at (11.5,-6.5) {\normalsize $2P$};
	
	\draw [->] (11,-4.5) -- (11,-2);
	
	\draw (11,-1.75) -- (11.5,-1.75) -- (11.75,-2) -- (10.75,-2) -- cycle;
	
	\draw [->] (12.5,-7) -- (12.5,-5);
	\draw [->] (12.7,-4.37) -- (12.7,-4) -- (11.5-.05,-4) -- (11.5-.05,-2);
	
	\draw[->] (12.7,-4) -- (12.7,0) -- (14,0);
	\fill (12.7,-4) circle [radius=.05];
	\node[right] at (12.7,-.3) {\Large codeword\_};
	\node[right] at (12.7,-.7) {\Large mem\_w\_data};
	
	\draw (12.85,-5.05) circle [radius=.05];
	\draw [->] (14,-6.5) -- (12.85,-6.5) -- (12.85,-5.1);
	
	\draw [->] (11.5+.1,-3.5) -- (11.5+.1,-2);
	\node at (11.7,-3.65) {\normalsize "0"};
	
	\draw (11.5+.1-.1,-2.75-.1) -- (11.5+.1+.1,-2.75+.1);
	\node[right] at (11.5+.1,-2.75) {\normalsize $2P$-1};
	
	\draw [->] (11.25,-1.75) -- (11.25,1);

	\node at (11.25,1.25) {\Large \texttt{w\_data}};

	\draw (11.25-.1,0.25-.1)--(11.25+.1,0.25+.1);
	\node[right] at (11.25,0.25) {\normalsize $2P$};
	
	\draw (.5,-8) -- (.5,-7) -- (13.5,-7) -- (13.5,-8);
	\node at (7,-7.75) {\Large PU};

	\node[and gate US, draw, logic gate inputs=nn, anchor=input 1,rotate=90] at ($(12.6,-5)$) (And1) {};

	\end{circuitikz}
	}
	\caption{$\beta$ memory interface circuit, where \texttt{r\_data\_0}, \texttt{r\_data\_1} and  \texttt{r\_data\_2} are the outputs of \texttt{bank0}, \texttt{bank1} and \texttt{bank2} respectively.}
	\label{fig:sez_mem_int:beta_interface}	
\end{figure}
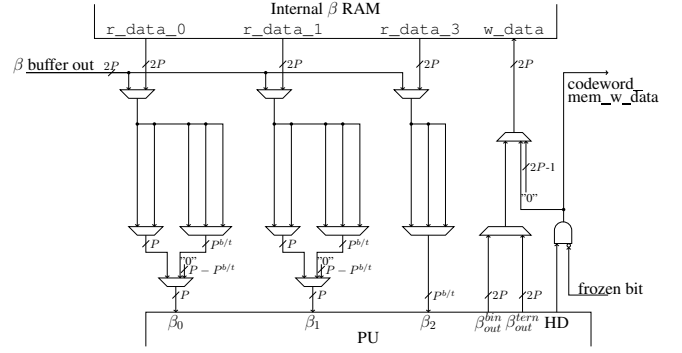

Figure \ref{fig:sez_mem_int:beta_interface} shows the interface architecture.
It is used both for reading and writing from the \textit{Internal $\beta$ RAM}:
\begin{itemize}
	\item Reading: operations involving $\beta$ values need either $P$ or $P^{b/t}$ bits per bank as input, while each word is composed of $2P$ bits. Thus, the relevant word parts are selected according to the actual number of elaborated LLRs for that node. 
	\item Writing: the data is selected between the CB results and the HD for the leaf nodes.
\end{itemize}

\subsection{Bypass registers}
Two bypass registers must be used since the memory system is RAM-based and, if a result is computed and ready to be stored at the $j$-th clock cycle, it can be correctly read only from the ($j+2$)-th cycle onwards, to avoid incurring conflicts.
So, for all the nodes at stage $s\le \log_2 2P$, bypass registers allow reading newly computed data already at the following clock cycle.
A $2QP$-bit register is used for the \textit{Internal LLR RAM}, while a second $2P$-bit register is necessary for the \textit{Internal $\beta$ RAM}.

\subsection{Control Unit}
The \textit{Control Unit} provides all the memory addresses to the memories and control signals to the datapath. It has been designed as several hierarchically controlled finite state machines. The decoding process follows the same approach of the tree exploration by means of different counters, which keep track of the status and of the number of visited leafs. The decoding process ends when a number of leafs equal to the code length has been visited.

\subsection{Multi-code support}
Memories are sized for a maximum code length $N_{max}$, but any code length $N\le N_{max}$, with $N$ a multiple of $2$ or $3$ is supported. Memory requirements are upper bounded by the largest combination of $\mathbf{T_2}$ kernels leading to $N_{max}$, since a higher number of stages are present in the decoding tree than in a mixed-kernel polar code with similar code length. The input code parameters allow to know when the leaf node stage is reached, and thus when the tree ascension has to start. The status counter in the CU uses foreknowledge of the number of kernels and their dimension to schedule the right operation at each stage: thus, any code rate and kernel order can be decoded without any change to the hardware. The total amount of bits required to store the code parameters for a code of length $N$ is $\left\lceil \log_2N\right\rceil + s_m \left( 2+2\left\lceil \log_2 \frac{N}{2P} \right\rceil \right)$, where $s_m$ is the number of kernel composing the code.
The PU has been designed independently of the code length.

\section{Implementation Results} \label{sec:impl}

The decoder architecture illustrated in the previous Section has been described in VHDL, verified with ModelSim, and synthesized with Cadence RTL Compiler on TSMC 65nm CMOS technology node.

\begin{figure}[t!]
	\centering
	\includegraphics[width=\columnwidth]{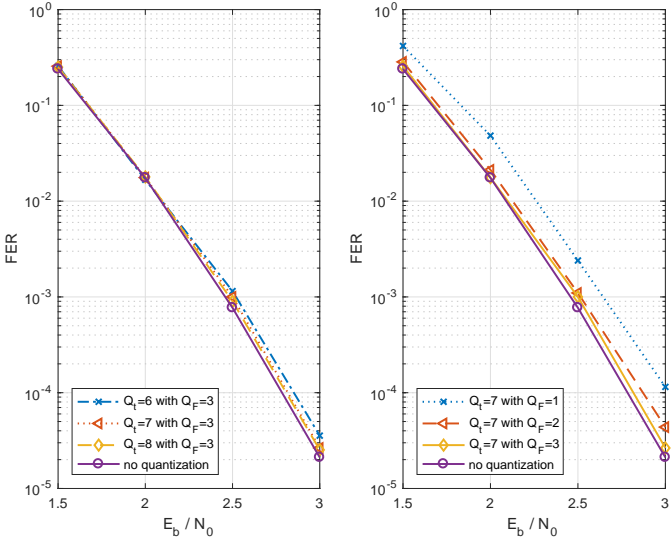}
	\caption{Error-correction performance of a $\mathcal{P}(4096,2048)$ with various $Q$ and $Q_f$ values.}
	\label{fig:sez_results:quant}	
\end{figure}

The choice of the number of LLR quantization bits $Q$ influences a substantial part of the computational hardware and memory width. In Figure \ref{fig:sez_results:quant} the error-correcting performance of a $\mathcal{P}(4096,2048)$ polar code is shown: between $Q=7$ and $Q=8$ curves there is not a significant difference, while choosing $Q=6$ leads to larger error figures with respect to floating point precision. Although the number of fractional bits $Q_f$ does not influence the hardware architecture, a high $Q_f$ requires a higher $Q$. In Figure \ref{fig:sez_results:quant} we can notice that $Q_f=3$ yields only minor FER degradation. Thus, for $N_{max}=4096$ we chose $Q=7$ and $Q_f=3$. Similar studies were performed in case of $N_{max}=1024$ and $N_{max}=256$, leading to $Q=6$, $Q_f=3$ in the first case and to $Q=5$, $Q_f=2$ in the second.

Table \ref{table:sez_results:performances} reports synthesis results for three sets of decoder parameters. Along with the parameters, the number of supported code lengths $\mathcal{N}$ and the maximum achievable frequency $f_{max}$ are shown. All implementations can run at more than one GHz. The $A^{\text{reg}}$ is the area occupation when all memories are synthesized as registers, while in $A^{\text{RAM}}$ all the memories are implemented as SRAM. For both estimations the logic and memory cells area percentages are shown.

\begin{table}[t!]
\setlength{\extrarowheight}{1.8pt}
	\caption{ASIC implementation results for TSMC 65nm CMOS technology, with number of supported code lengths $\mathcal{N}$, maximum frequency $f_{max}$, register-based area occupation $A^{\text{reg}}$ and RAM-based area occupation $A^{\text{RAM}}$.}
	\label{table:sez_results:performances}
	\begin{center}
		\begin{tabular}{cc|c|c|c}
			\toprule
			$N_{max}$&&$4096$&$1024$&$256$\\
			$P$&&$120$&$60$&$18$\\
			$Q$&&$7$&$6$&$5$\\
			$\mathcal{N}$&&55&40&27\\
			$f_{max}$ &[GHz]&1.06&1.11&1.23\\
			\midrule
			$A^{\text{reg}}$ &[mm$^2$]&2.63&0.62&0.14\\
			Combinational &[\%]&45.0 &38.9 &40.3 \\
			Sequential &[\%]&55.0&61.1& 59.7\\
			\midrule
			$A^{\text{RAM}}$ &[mm$^2$]&2.01&0.46&0.11\\
			Combinational &[\%]&58.9 & 56.7 &55.2\\
			Registers &[\%]&28.6 &28.9 &31.2 \\
			RAM &[\%]&12.5&14.4&13.6\\ 

\bottomrule
		\end{tabular}
	\end{center}
\end{table}
The latency of the decoding phase depends on the number $P$ of PEs, on the number of kernels $s_m$, on the kernels dimension and their order. 

The decoding latency, measured in clock cycles (CCs) can be computed as:
\begin{equation}
\displaystyle \mathcal{L}= \sum_{s=1}^{s_m}  \left\lceil \frac{N_s}{2P} \right\rceil  \left((n_s+1) \frac{N}{N_s} -1\right)~.
\label{eqn:sez_results:latency}
\end{equation}

In Table \ref{table:sez_results:latency} some polar code timing performance are shown, where $\mathcal{L}$ is the decoding latency, $f$ is the achievable frequency, and $T$ the coded throughput. They consider a wide range of code parameters over three different decoder implementations. Since the kernel order impacts the decoding latency, dimension of each kernel has been reported, from left to right as in the Kronecker product. It is possible to see that the achievable frequency is consistently above 1 GHz, and that the coded throughput ranges from $350$ to $615$ Mbps.
\begin{table}[t!]
\setlength{\extrarowheight}{1.8pt}
\caption{Latency $\mathcal{L}$ and coded throughput $T$ of various polar codes with three different decoder implementations.}
\label{table:sez_results:latency}
\begin{center}
	{\scriptsize
	\begin{tabular}{c|cc|cccc}
		\toprule
		\textsc{code}&	\multicolumn{2}{c|}{\textsc{decoder}} &$\mathcal{L}$& $f$ &$T$&$T$\\
		\textsc{parameters} & 	\multicolumn{2}{c|}{\textsc{parameters}}  &[CCs]&[GHz]&[bpc] &[Mbps]\\
		\hline \hline
		\{2,3,2,2,2,3,3,3,3\} &\multicolumn{2}{c|}{$N_{max}=4096$}& \multirow{2}{*}{7965}& \multirow{2}{*}{1.06}&\multirow{2}{*}{0.49}&\multirow{2}{*}{519.4}\\
		$N=3888$ & $P$=120 & $Q$=7&&&\\
		\hline
		\{2,3,3,2,3,3,3,3\} &\multicolumn{2}{c|}{$N_{max}=4096$}& \multirow{2}{*}{5953}& \multirow{2}{*}{1.06}&\multirow{2}{*}{0.49}&\multirow{2}{*}{519.4}\\
		$N=2916$ & $P$=120 & $Q$=7&&&\\
		\hline
		\{2,2,2,2,2,2,3,3,3\} &\multicolumn{2}{c|}{$N_{max}=4096$}& \multirow{2}{*}{3548}& \multirow{2}{*}{1.06}&\multirow{2}{*}{0.49}&\multirow{2}{*}{519.4}\\
		$N=1728$ & $P$=120 & $Q$=7&&&\\
		\hline
		\{3,2,2,2,2,2,2,2,2,2\} &\multicolumn{2}{c|}{$N_{max}=4096$}& \multirow{2}{*}{4663}& \multirow{2}{*}{1.06}&\multirow{2}{*}{0.33}&\multirow{2}{*}{350.6}\\
		$N=1536$ & $P$=120 & $Q$=7&&&\\
		\hline
		\{2,2,3,2,2,2,2,2,2\}& \multicolumn{2}{c|}{$N_{max}=1024$}&\multirow{2}{*}{2326}& \multirow{2}{*}{1.11}&\multirow{2}{*}{0.33}&\multirow{2}{*}{366.5}\\
		$N=768$ & $P$=60& $Q$=6&&&\\
		\hline
		\{2,2,2,2,2,2,3,3\}& \multicolumn{2}{c|}{$N_{max}=1024$}&\multirow{2}{*}{1234}& \multirow{2}{*}{1.11}&\multirow{2}{*}{0.47}&\multirow{2}{*}{521.7}\\
		$N=576$ & $P$=60& $Q$=6&&&\\
		\hline
		\{3,2,2,2,2,2,2,2\}& \multicolumn{2}{c|}{$N_{max}=1024$}&\multirow{2}{*}{1156}& \multirow{2}{*}{1.11}&\multirow{2}{*}{0.33}&\multirow{2}{*}{368.7}\\
		$N=384$ & $P$=60& $Q$=6&&&\\
		\hline
		\{2,2,3,3,3,3\}& \multicolumn{2}{c|}{$N_{max}=1024$}&\multirow{2}{*}{652}& \multirow{2}{*}{1.11}&\multirow{2}{*}{0.50}&\multirow{2}{*}{555.0}\\
		$N=324$ & $P$=60& $Q$=6&&&\\
		\hline
		\{3,3,3,3,3\} & \multicolumn{2}{c|}{$N_{max}=256$}&\multirow{2}{*}{519}& \multirow{2}{*}{1.23}&\multirow{2}{*}{0.47}&\multirow{2}{*}{578.1}\\
		$N=243$&$P$=18 & $Q$=5&&&\\
		\hline
		\{3,2,2,2,2,2,2\} & \multicolumn{2}{c|}{$N_{max}=256$}&\multirow{2}{*}{587}& \multirow{2}{*}{1.23}&\multirow{2}{*}{0.32}&\multirow{2}{*}{402.3}\\
		$N=192$&$P$=18 & $Q$=5&&&\\
		\hline		
		\{2,2,2,3,2,2\} &\multicolumn{2}{c|}{$N_{max}=256$}&\multirow{2}{*}{272}& \multirow{2}{*}{1.23}&\multirow{2}{*}{0.35}&\multirow{2}{*}{434.1}\\
		$N=96$&$P$=18 & $Q$=5&&&\\
		\hline
		\{3,3,3,3\} &\multicolumn{2}{c|}{$N_{max}=256$}&\multirow{2}{*}{162}& \multirow{2}{*}{1.23}&\multirow{2}{*}{0.50}&\multirow{2}{*}{615.0}\\
		$N=81$&$P$=18 & $Q$=5&&&\\
		\hline
		\{3,2,2,2,2\} & \multicolumn{2}{c|}{$N_{max}=256$}&\multirow{2}{*}{137}& \multirow{2}{*}{1.23}&\multirow{2}{*}{0.35}&\multirow{2}{*}{430.9}\\
		$N=48$ & $P$=18 & $Q$=5&&&\\
		\bottomrule
	\end{tabular}}
\end{center}
\end{table}

In Table \ref{table:sez_results:comparison} the implementation results of the proposed decoder have been compared to rate-flexible purely binary decoders in the state of the art, since to the best of our knowledge this is the first multi-kernel decoder in literature. All decoders have been implemented with 65~nm CMOS technology, and target a code with $N=1024$, that for our work corresponds to $N_{max}$ as well.
Both in \cite{leroux} and \cite{Fan_Tsui} semi-parallel architectures are proposed, supporting the SC algorithm and a single fixed code length. The reported results for \cite{Fan_Tsui} refer to their best devised architecture, called folded high performance partial sum network. It limits the number of processing elements by folding highly parallel operations and performing them in several clock cycles, thus increasing hardware utilization. Observing the bit-per-cycle (bpc) throughput in Table \ref{table:sez_results:comparison}, it can be noticed that both \cite{leroux} and \cite{Fan_Tsui} outperform the proposed decoder for the considered purely binary codes. The reason can be found in the additional clock cycles required for $comb$ operations in our architecture: since different kernel orders are supported, the sequence of (\ref{eqn:sez_prel:comb_bin}) and (\ref{eqn:sez_prel:comb_tern}) is not always the same. Thus, it is not possible to hardwire an XOR tree to compute the $comb$ at all stages in one clock cycle, like in decoders supporting only binary kernels: separate clock cycles are spent to perform the $comb$ operations according to the correct kernel order. On the other hand, \cite{leroux} and \cite{Fan_Tsui} consider only binary kernels and, implementing a tree of $comb$ operations and eventually selecting a partial result, $\beta$ values are computed in the same clock cycle immediately after the $g$. This is not affecting the critical path in a significant way since only few XOR gates are added. As shown in Table \ref{table:sez_results:latency}, codes constructed with higher-dimension kernels yield a higher throughput. When decoding a ternary node, due to the higher utilization factor of the PEs and the higher number of useful computations in each clock cycle, the number of clock cycles needed to decode a codeword is lower. Moreover, latency-reduction techniques like the ones presented in \cite{Zhang_ICC12,yuan_2bitDecoding} can be easily adapted to the proposed architecture.
 
The proposed decoder yields a higher area occupation than both \cite{leroux} and \cite{Fan_Tsui}. This is mainly due to the higher quantization parameter $Q$ and to the support to ternary functions. Mixed PEs require $\times2.57$ LUTs on FPGA and $\times2.10$ area occupation with respect to the purely binary ones. However, our decoder is completely code-length flexible and supports multiple kernel sizes, any code rate and any kernel order. Moreover, it can achieve the highest frequency among the considered works, and a higher throughput in Mbps than \cite{leroux}.

Semi-parallel SC-based decoders in literature, while supporting only binary kernels and often being designed targeting a single code, share the basic multi-PE structure of our work. For the sake of completeness, in Table \ref{table:sez_results:comparison} we consider also \cite{Dizdar_TCASI} and \cite{giard_unrolled}. These architectures are very different from semi-parallel decoders, but guarantee a certain degree of flexibility. The decoder in \cite{giard_unrolled} can decode a fixed set of combinations of code lengths and code rates, while the architecture proposed in \cite{Dizdar_TCASI} is rate-flexible. Both architectures are able to achieve a higher throughput than the proposed decoder, at the cost of larger area occupation and a lower degree of flexibility.

\begin{table}[t!]
\setlength{\extrarowheight}{1.8pt}
\caption{Comparison with the state of the art, $N=1024$ polar codes, coded throughput $T$, area $A$.}
	\label{table:sez_results:comparison}
	\begin{center}
		\begin{tabular}{ccccccc}
		\toprule
		\multirow{2}{*}{\textsc{Decoder}}& \multirow{2}{*}{$P$}& \multirow{2}{*}{$Q$}&$f$&$T$&$T$ & $A$\\
		& &&[GHz]&[bpc]& &[mm$^2$]\\
		\midrule
		This work&$60$& $6$  &1.11&0.33&361.98 Mbps&0.46\\	
		\cite{leroux} &  $64$ & $5$  &0.50&0.49& 246.10 Mbps & 0.31\\
		\cite{Fan_Tsui} & $64$& $5$  &1.01&0.49& 497.28 Mbps  & 0.07\\
		\midrule
		\cite{Dizdar_TCASI} & -- & $5$ & 0.0025 & 1418&3.54 Gbps  & 1.68\\ 
		\cite{giard_unrolled} & -- & $5$ & 0.65 & 39.4& 25.60 Gbps & 1.44\\
		\bottomrule
		\end{tabular}
	\end{center}
\end{table}

\section{Conclusion} \label{sec:conc}
In this work, we have proposed the first polar code decoder architecture supporting kernels of different sizes. It implements the successive cancellation algorithm, and can support any code rate, any sequence of binary and ternary kernels and any code length $N\le N_{max}$ that can be expressed as a combination of binary and ternary kernels. The decoder can achieve a frequency of more than a GHz in 65 nm CMOS technology, and a throughput of $615$ Mb/s. The area occupation ranges between $0.11$~mm$^2$ for $N_{max}=256$ and $2.01$~mm$^2$ for $N_{max}=4096$. Implementation results show an unprecedented degree of flexibility: with $N_{max}=4096$, up to 55 code lengths can be decoded with the same hardware, along with any kernel sequence and code rate. 


\bibliographystyle{IEEEtran}

\end{document}